\documentclass[journal]{IEEEtran}
  \usepackage[nocompress]{cite}
  \usepackage{cite}
  \usepackage{subfig}
  \usepackage{graphicx}
  \usepackage[cmex10]{amsmath}
  \usepackage{epstopdf}
  \usepackage{algorithm, longtable}
  \usepackage[noend]{algpseudocode}
  \usepackage{multirow}
  \usepackage[table,xcdraw]{xcolor}

\usepackage[acronym,nopostdot]{glossaries}

\makeglossaries 
\newacronym{leo}{LEO}{Low Earth Orbit}
\newacronym{satnets}{SatNet}{Satellite Network}
\newacronym{3gpp}{3GPP}{3rd Generation Partnership Project}
\newacronym{nr}{NR}{New Radio}
\newacronym{qos}{QoS}{Quality-of-Service}
\newacronym{meo}{MEO}{Medium Earth Orbit}
\newacronym{geo}{GEO}{Geosynchronous Earth Orbit}
\newacronym{iot}{IoT}{ Internet  of  Things}
\newacronym{m2m}{M2M}{ Machine-to-Machine}
\newacronym{ntn}{NTN}{Non-Terrestrial Network}
\newacronym{haps}{HAPSs}{High  Altitude  Platform  Systems}
\newacronym{uav}{UAV}{Unmanned Aerial Vehicle}
\newacronym{5g}{5G}{Fifth Generation}
\newacronym{5g+}{5G+}{Fifth Generation and Beyond}
\newacronym{6g}{6G}{Sixth Generation}
\newacronym{ue}{UE}{User Equipment}
\newacronym{ngso}{NGSO}{Non-Geostationary Orbiting}
\newacronym{heo}{HEO}{Highly Eccentric Orbiting}
\newacronym{si}{SI}{Study Item}
\newacronym{wi}{WI}{Work Item}
\newacronym{tsg}{TSG}{Technical Specification Group}
\newacronym{ran}{RAN}{Radio Access Network}
\newacronym{sa}{SA}{Service and System Aspects}
\newacronym{ct}{CT}{Core Network and Terminals}
\newacronym{embb}{eMBB}{Enhanced Mobile Broadband}
\newacronym{mimo}{MIMO}{Multiple-Input Multiple-Output}
\newacronym{iab}{IAB}{Integrated Access Backhaul}
\newacronym{xr}{XR}{Extended Reality}
\newacronym{ai}{AI}{Artificial Intelligence}
\newacronym{ml}{ML}{Machine Learning}
\newacronym{ca}{CA}{Carrier Aggregation}
\newacronym{dc}{DC}{Dual-Connectivity}
\newacronym{mrmc}{MR-MC}{Multi-Radio/Multi-Connectivity}
\newacronym{ris}{RIS}{Reconfigurable Intelligent Surfaces}
\newacronym{iiot}{IIoT}{Industrial Internet of Things}
\newacronym{urllc}{URLLC}{Ultra-Reliable Low-Latency Communication}
\newacronym{son}{SONs}{Self-Organizing Networks}
\newacronym{mdt}{MDT}{Minimization of Drive Test}
\newacronym{vsat}{VSATs}{Very Small Aperture Terminals}
\newacronym{isl}{ISL}{Inter-Satellite Links}
\newacronym{ul}{UL}{Uplink}
\newacronym{dl}{DL}{Downlink}
\newacronym{ici}{ICI}{Inter-Carrier Interference}
\newacronym{sps}{SPS}{Semi-Persistent Scheduling}
\newacronym{tti}{TTI}{Transmission Time Interval}
\newacronym{snr}{SNR}{Signal-to-Noise Ratio}
\newacronym{harq}{HARQ}{Hybrid Automatic Repeat Request}
\newacronym{rtt}{RTT}{Round Trip Time}
\newacronym{ta}{TA}{Tracking Area}
\newacronym{rlc}{RLC}{Radio Link Control}
\newacronym{arq}{ARQ}{Automatic Repeat Request}
\newacronym{acm}{ACM}{Automatic Coding and Modulation}
\newacronym{rar}{RAR}{Random Access Response}
\newacronym{prach}{PRACH}{Physical Random Access Channel}
\newacronym{tdd}{TDD}{Time Division Duplex}
\newacronym{fdd}{FDD}{Frequency Division Duplex}
\newacronym{ptrs}{PT-RS}{Phase Tracking Reference Signal}
\newacronym{papr}{PAPR}{Peak-to-Average Power Ratio}
\newacronym{cpofdm}{CP-OFDM}{Cyclic Prefix – Orthogonal Frequency Division Multiplexing}
\newacronym{itu}{ITU}{International Telecommunication Union}
\newacronym{rat}{RAT}{Radio Access Technology}
\newacronym{wlan}{WLAN}{Wireless Local Area Network}
\newacronym{fss}{FSS}{Fixed Satellite Service}
\newacronym{mss}{MSS}{Mobile Satellite Service}
\newacronym{bss}{BSS}{Broadcast Satellite Service}
\newacronym{ip}{IP}{Internet Protocol}
\newacronym{wrc}{WRC}{World Radio Conference}
\newacronym{esim}{ESIM}{Embedded SIM }
\newacronym{ietf}{IETF}{Internet Engineering Task Force}
\newacronym{tcp}{TCP}{Transmission Control Protocol}
\newacronym{udp}{UDP}{User Datagram Protocol}
\newacronym{sctp}{SCTP}{Stream Control Transmission Protocol}
\newacronym{mptcp}{MPTCP}{Multi-Path TCP}
\newacronym{etsi}{ETSI}{European Telecommunications Standards Institute}
\newacronym{mec}{MEC}{Mobile Edge Computing}
\newacronym{nfv}{NFV}{Network Function Virtualization}
\newacronym{nin}{NIN}{Non-IP Networking}
\newacronym{mwt}{mWT}{Millimeter Wave Transmission}
\newacronym{mano}{MANO}{Management and Orchestration}
\newacronym{osm}{OSM}{Open Source MANO}
\newacronym{5gppp}{5G PPP}{5G Infrastructure Public Private Partnership}
\newacronym{ecc}{ECC}{Electronic Communication Committee}
\newacronym{nict}{NICT}{National Institute of Information and Communication Technology}
\newacronym{ingr}{INGR}{International Network Generation Roadmap}
\newacronym{qoe}{QoE}{Quality of Experience}
\newacronym{dvb}{DVB}{Digital Video Broadcasting}
\newacronym{ccsds}{CCSDS}{Consultative Committee for Space Data Systems}
\newacronym{aeec}{AEEC}{Airlines Electronic Engineering Committee}
\newacronym{mip}{MIPv6}{Mobile Internet Protocol version 6}
\newacronym{pmip}{PMIPv6}{Proxy Mobile Internet Protocol version 6}
\newacronym{sdn}{SDN}{Software Defined Network}
\newacronym{vnf}{VNF}{Virtualized Network Function}
\newacronym{isg}{ISG}{Industry Specification Group}
\newacronym{eni}{ENI}{Experiential Networked Intelligence}
\newacronym{sen}{SEN}{Self-Evolving Network}

\makeatletter
\def\BState{\State\hskip-\ALG@thistlm}
\makeatother

\begin{document}

\title{\textcolor{black}{LEO} Satellites in 5G and Beyond Networks: A Review from a Standardization Perspective}

\author{Tasneem Darwish,                               ~\IEEEmembership{Senior Member,~IEEE,}         Gunes Karabulut Kurt, 
        ~\IEEEmembership{Senior Member,~IEEE,} Halim Yanikomeroglu, ~\IEEEmembership{Fellow, ~IEEE,}  Michel Bellemare, and Guillaume Lamontagne 
\thanks{Tasneem Darwish and Halim Yanikomeroglu are with the Department of Systems and Computer Engineering, Carleton University, Ottawa, Canada (e-mail: tasneemdarwish@sce.carleton.ca; halim@sce.carleton.ca).\protect\\
Gunes Karabulut Kurt is with the Poly-Grames Research Center, Department of Electrical Engineering,  Polytechnique Montr\'eal, Montr\'eal, Canada (e-mail: gunes.kurt@polymtl.ca). \protect\\
Michel Bellemare and Guillaume Lamontagne are with the Division of Satellite Systems, MDA, Canada (e-mail: Guillaume.Lamontagne@mda.space; Michel.Bellemare@mda.space).

}
}


\IEEEtitleabstractindextext{%
\begin{abstract}
\acrfull{leo} \acrfull{satnets} with their mega-constellations are expected to play a key role in providing ubiquitous Internet and communications services in the future. LEO SatNets will provide wide-area coverage and support service availability, continuity, and scalability. To support the integration of SatNets and terrestrial \acrfull{5g} networks and beyond,  the satellite communication industry has become increasingly involved with the  \acrfull{3gpp} standardization activities for 5G. In this work, we review the 3GPP standardization activities for the integration of SatNets in 5G and beyond. The 3GPP use cases of SatNets are highlighted and potential requirements to realize them are summarized as well. The impacted areas of \acrfull{nr} are discussed with some potential solutions. The \textcolor{black}{foreseen} requirements for the management and orchestration of SatNets within 5G are described. Future standardization directions  are discussed to support the full integration of SatNets in \acrfull{6g} with the goal of ubiquitous global connectivity.

\end{abstract}

~~~\\}


\maketitle

\IEEEdisplaynontitleabstractindextext

\IEEEpeerreviewmaketitle

\glssetwidest{HAPS-SMBS}
\hspace{1cm}\printglossary[style=alttree,type=\acronymtype,title=\textbf{Abbreviations},nogroupskip, nonumberlist]

~\\

~\\
\IEEEraisesectionheading{\section{Introduction}\label{s1}}


The exponential growth of the number of connected smart devices, the ever-increasing demand for new services with stringent \acrfull{qos} requirements, and the need for continuous connectivity everywhere are creating difficult-to-meet challenges for the terrestrial telecommunication sector. Ericsson Mobility Report \cite{ericsson2020} estimated that there will be 3.5 billion 5G subscriptions by the end of 2026. Moreover, smart cities, intelligent transportation systems, and automated industrial sites will involve billions of sensors and devices that will create a huge traffic load on terrestrial communication networks and that will require continuous coverage with efficient mobility support. In this context, satisfying all user requests and providing the desired QoS anytime and anywhere—even when traveling on cruises, high-speed trains, and airplanes—are two of the main challenges for future telecommunication systems.

Satellite communication networks utilize spaceborne platforms which include \acrfull{leo} satellites, \acrfull{meo} satellites, and \acrfull{geo} satellites. Over the past several years, the world has witnessed a resurging interest in broadband provisioned by LEO satellite networks (SatNets) with large satellite constellations (e.g., Starlink, Kuiper, OneWeb, and \textcolor{black}{Lightspeed}). With their capacity to form networks among satellites, LEO SatNets will play a significant role in future integrated networks. This new satellite architecture will revolutionize traditional communication networks with its promising benefits of  service continuity, wide-area coverage, and availability  for critical communications and emerging applications (e.g., \acrfull{iot} devices/\acrfull{m2m}, and intelligent transportation systems), and enabling network scalability.

In the history of telecommunications, satellites and terrestrial networks have always been considered two independent ecosystems, and their standardization efforts have proceeded independently of each other. To benefit from the market potential of integrating satellite networks into the 5G ecosystem, there has been a growing interest from the satellite communication industry to participate in the 3GPP standardization effort for 5G. 3GPP classifies satellites as part of the \acrfull{ntn}, which is considered as a complement to the terrestrial networks. As defined by 3GPP, an NTN is a network where spaceborne platforms (i.e., GEO, MEO, LEO satellites) or airborne platforms (i.e., \acrfull{haps} \cite{kurt2021vision}) act either as a relay node or as a base station.  However, the focus of 3GPP NTN work has been on satellites for communication, whereas HAPSs are considered a special case of a satellite system. Although \acrfull{uav} constitute a part of airborne networks, their line of standardization has been carried out in a separate track in 3GPP. In this survey the focus is on satellites in the context of 5G and beyond (i.e., 5G+).

In Release 14, 3GPP started to consider satellite communications in a study on scenarios and requirements for next generation access technologies \cite{3gpp38913}. In subsequent releases (i.e., Releases 15, 16, and 17), 3GPP considered satellite communication networks from several aspects, such as \acrfull{nr}, architectures, use cases, scenarios, management, and orchestration. This survey comprehensively reviews the works on the standardization of satellite communication networks with a focus on 3GPP activities.

\begin{table*}[h]
\centering
\caption{Comparison of existing survey papers.}
\label{TableComparison}
\begin{scriptsize}
\begin{tabular}{|p{1.5cm}|p{1.5cm}|p{1.5cm}|p{1.5cm}|p{1.5cm}|p{1.5cm}|p{1.5cm}|p{1.5cm}|}
\hline
\rowcolor[HTML]{CBCEFB} 
Reference & Discussion of 3GPP activities Releases 15, 16, 17 & Discussion of 3GPP activities Release 18 & Summary of other standardization organization activities & Impacted areas of NR in integrated 5G SatNet & Management Orchestration of SatNet & 3GPP Satellite Use cases & Architectures \\ \hline

\cellcolor[HTML]{EFEFEF} \cite{rinaldi2020non} &  Focused on 15 \& 16  & \cellcolor[HTML]{FFCCC9} No & \cellcolor[HTML]{FFCCC9} No & \cellcolor[HTML]{FFCCC9} No & \cellcolor[HTML]{FFCCC9} No & \cellcolor[HTML]{FFCCC9} No & \cellcolor[HTML]{B8E8F6} Yes \\ \hline

\cellcolor[HTML]{EFEFEF} \cite{burleigh2019connectivity}& \cellcolor[HTML]{FFCCC9} No & \cellcolor[HTML]{FFCCC9} No & \cellcolor[HTML]{FFCCC9} No & \cellcolor[HTML]{FFCCC9} No & \cellcolor[HTML]{FFCCC9} No & \cellcolor[HTML]{FFCCC9} No & \cellcolor[HTML]{FFCCC9} No \\ \hline

\cellcolor[HTML]{EFEFEF} \cite{guidotti2020architectures} & \cellcolor[HTML]{B8E8F6} Yes & \cellcolor[HTML]{FFCCC9}  No & \cellcolor[HTML]{FFCCC9} No & Focused on three areas only & \cellcolor[HTML]{FFCCC9} No  & \cellcolor[HTML]{FFCCC9} No & \cellcolor[HTML]{B8E8F6} Yes   \\ \hline

\cellcolor[HTML]{EFEFEF} \cite{guidotti2019architectures}& \cellcolor[HTML]{FFCCC9} No & \cellcolor[HTML]{FFCCC9} No & \cellcolor[HTML]{FFCCC9} No  & Focused on two areas only & \cellcolor[HTML]{FFCCC9} No & \cellcolor[HTML]{FFCCC9} No & \cellcolor[HTML]{B8E8F6} Yes \\ \hline

\cellcolor[HTML]{EFEFEF} \cite{lin20215g} & \cellcolor[HTML]{B8E8F6} Yes  & \cellcolor[HTML]{FFCCC9} No & \cellcolor[HTML]{FFCCC9} No  & \cellcolor[HTML]{FFCCC9} No  & \cellcolor[HTML]{FFCCC9} No & \cellcolor[HTML]{FFCCC9} No & \cellcolor[HTML]{B8E8F6} Yes   \\ \hline

\cellcolor[HTML]{EFEFEF} \cite{kodheli2020} & Very brief  & \cellcolor[HTML]{FFCCC9} No & \cellcolor[HTML]{FFCCC9} No  & \cellcolor[HTML]{FFCCC9} No  & \cellcolor[HTML]{FFCCC9} No & \cellcolor[HTML]{B8E8F6} Yes & \cellcolor[HTML]{B8E8F6} Yes   \\ \hline

\cellcolor[HTML]{EFEFEF} Our work & \cellcolor[HTML]{B8E8F6} Yes & \cellcolor[HTML]{B8E8F6} Yes & \cellcolor[HTML]{B8E8F6} Yes  & \cellcolor[HTML]{B8E8F6} Yes & \cellcolor[HTML]{B8E8F6} Yes & \cellcolor[HTML]{B8E8F6} Yes & \cellcolor[HTML]{B8E8F6} Yes   \\ \hline

\end{tabular}
\end{scriptsize}
\end{table*}

\subsection{Motivation and contributions}

In the literature, there are several reviews and surveys on satellite systems. Table \ref{TableComparison} provides a comparative overview of the existing surveys that discuss standardization efforts in the area of SatNets. The most common topic among existing surveys is the different 3GPP architectures for 5G with satellite access. Nevertheless, the discussion on SatNet related 3GPP  work therein stays at a high level as exiting surveys objective is to give a broad view of the SatNet state-of-the-art.  In contrast, this article aims to offer a dedicated and comprehensive review of the 3GPP standardization work in the area of SatNets. The scope of this review spans across multiple areas, from radio access network to services and system aspects. Regarding the contributions, this work:

\begin{itemize}
    \item Provides a comprehensive survey of the 3GPP standardization activities in the area of satellite networks and communication from Release 14 to Release 18.
    \item Highlights the 3GPP use cases for satellite access in 5G and their applications.
    \item Discusses the required adaptation of NR for SatNets from the 3GPP perspective. 
    \item Summarizes the potential requirements for the management and orchestration of integrated satellite components in a 5G network.
    \item Presents an overview of standardization efforts from organizations other than the 3GPP.
    \item Discusses future directions to be taken in standardization efforts for 6G satellite communication networks.
\end{itemize}

\subsection{Paper organization}
The remainder of this paper is organized as follows. Section~\ref{sec2} gives a brief description of the satellite access network elements and highlights the characteristics of satellite networks from the perspective of 3GPP. Section \ref{sec3} gives an overview of 3GPP standardization activities with respect to satellite networks. Section \ref{sec32} describes the satellite access networks use cases in the context of 5G. The architectures of integrated satellite access networks and 5G networks are presented in Section \ref{sec4}. Section \ref{sec5} discusses the adaptation of the New Radio for satellite networks. In particular, subsection \ref{sec51} specifies the constraints associated with satellite networks and subsection \ref{sec52} highlights the NR impacted areas and their potential solutions. Section \ref{sec6} sheds the light on the management and orchestration aspects of 5G networks with integrated satellite access components. The activities of the non-3GPP standardization organizations are discussed in Section \ref{sec7}. Section \ref{sec8} highlights important standardization directions that are required for the full integration of satellite networks with 6G networks. Section \ref{sec9} draws the essential conclusions.

\section{Satellite Access Network Elements and Characteristics in 3GPP Standardization}\label{sec2}

Due to the wide service coverage of SatNets and their reduced vulnerability to physical attacks and natural disasters, the 3GPP sought to define the expected role of SatNets in 5G+ through the following points in TR 38.811 \cite{3gpp38811}:

\begin{itemize}
    \item Provide 5G service in unserved areas that cannot be covered by terrestrial 5G networks (e.g., isolated and remote areas, on aircrafts and ships) and underserved areas (e.g., suburban and rural areas).
    \item Upgrade the performance of limited terrestrial networks in a cost-effective manner.
    \item Support the reliability of 5G service by providing service continuity for M2M/IoT devices or for passengers on-board moving platforms and ensuring service availability anywhere, especially for critical communications and railway, maritime, and aeronautical communications.
    \item Enable 5G network scalability by providing efficient multicast and broadcast resources for data delivery towards the network edges or even user terminal.
\end{itemize}

As mentioned in 3GPP TR 38.811, satellite access networks consist of the following elements \cite{3gpp38811}:
\begin{itemize}
    \item \acrfull{ue} or a specific terminal to the satellite system in case the satellite doesn't serve UEs directly.
    \item A service link which is the radio link between the UE and the space platform.
    \item A space platform carrying a payload which may have one of these two configurations:
    \begin{itemize}
        \item A bent-pipe payload that performs radio frequency filtering, frequency conversion, and amplification.
        \item A regenerative payload offering radio frequency filtering, frequency conversion, and amplification as well as demodulation and decoding, switch and/or routing, coding and/or modulation. It is equivalent to how a base station functions (e.g., gNB) on-board a satellite. 
    \end{itemize}
    \item Inter-satellite links (ISLs) in case of a regenerative payload and a constellation of satellites. An \acrshort{isl} may operate in RF frequency or optical bands.
    \item Gateways that connect the satellite access network to the core network. 
    \item Feeder links which refer to the radio links between the gateways and the space platform.
\end{itemize}

In 3GPP TR 22.822 \cite{3gpp22822} the orbits that are used for communication satellites are calssified into the following four types:
\begin{itemize}
    \item GEO satellites, located precisely along the plane of the Equator at an altitude of 35,786 km. GEOs  orbit at the same rate as the earth's rotation. Therefore, a GEO satellite can provide continuous coverage.
    \item \acrfull{ngso} satellites which do not stand still with respect to the earth. To provide service continuity over time, a number of satellites (a constellation) is required to meet this requirement; the lower the altitude, the higher the number of satellites. Different types of NGSO satellites are listed below:
\begin{itemize}
    \item LEO satellites, with altitudes ranging from 500 km to 2,000 km, and with inclination angles of the orbital plane ranging from 0 \textcolor{black}{up to 180 degrees (prograde and retrograde orbits)}. These constellations are situated above the International Space Station and debris, and below the first Van Allen belt.
    \item MEO satellites, with altitudes ranging from 8,000 to 20,000 km. The inclination angles of the orbital plane range from 0 \textcolor{black}{up to 180 degrees (prograde and retrograde orbits)}. These constellations are situated above the Van Allen belts.
    \item \acrfull{heo} satellites, with a range of operational altitudes between 7,000 km and more than 45,000 km. The inclination angle is selected so as to compensate, completely or partially, for the relative motion of the earth with respect to the orbital plane, allowing the satellite to cover successively different parts of northern land masses (e.g. Western Europe, North America, and Northern Asia).
\end{itemize}
\end{itemize}


\section{Overview of Satellite Access Network 3GPP Standardization  and Use Cases} \label{sec3}
From the early stage of 5G standardization activities, 3GPP considered the integration of satellite systems to be a valuable asset to complement and integrate terrestrial NR networks. Therefore, several Study Items (SIs) and Work Items (WIs) were initiated within the \acrfull{tsg} of the \acrfull{ran}, the \acrfull{sa}, and the \acrfull{ct}. 

As an extension to terrestrial networks, satellites were first mentioned in a deployment scenario of 5G in 3GPP TR 38.913 Release 14. This was to provide 5G communication services for areas where  terrestrial coverage was not available and also to support services that could be accessed more efficiently through satellite systems, such as broadcasting services and delay-tolerant services. In 3GPP TS 22.261 Release 15, ``Service requirements for next generation new services and markets," the first analyses that described the significant role that satellites could play in 5G systems were provided. However, the focus of TS 22.261 was on  satellites in NR for industrial and mission critical  services. 

The increasing interest in integrating satellites with 5G led to the definition of NTNs by 3GPP. Although NTNs include other aerial systems (e.g., HAPSs), the 3GPP community considers satellites as the main case and other aerial systems as a special case of satellites. Release 16 is frozen and  Release 17 is still open with an expected deadline on December 2021. Release 17 is working on NTNs for 5G systems, which adopt satellites to support underserved areas (e.g., isolated and remote areas, onboard aircrafts and vessels). 



\subsection{3GPP standardization activities for satellite access networks} \label{sec31}
Through Releases 15, 16, 17, and 18, 3GPP launched several standardization activities to support the integration of SatNets and 5G terrestrial networks. The following points give an overview of the satellite-related 3GPP standardization activities:

\begin{itemize}
    \item Release 15: In 2017, two SIs were initiated: (1) 3GPP TR 38.811 ``Study on NR to support Non-Terrestrial Networks" under the RAN TSG; and (2) 3GPP TR 22.822 ``Study on using Satellite Access in 5G" under the SA TSG. The first SI aimed to define the NTN deployment scenarios and their system parameters as well as to identify the required NR adaptation to accommodate NTNs. Also preliminary solutions were introduced to address the impacted areas of NR. Although the second \acrshort{si} was initiated in 2017, it was moved to Release 16.
    
    \item Release 16: The SA TSG  had four activities: (1) an SI on ``Study on using satellite access in 5G" \cite{3gpp22822}; (2) an SI on ``Study on architecture aspects for using satellite access in 5G" \cite{3gpp23737}; (3) a \acrshort{wi} on ``Integration of Satellite Access in 5G" (WI\#800010-5GSAT, Release 16); and (4) an SI on ``Study on management and orchestration aspects with integrated satellite components in a 5G network" \cite{3gpp28808}. The first and second SIs introduced a number of use cases for the provision of services in the integrated 5G and satellite-based access components. This led to identifying corresponding modified or new requirements related to connectivity, roaming, QoS, UE, security, and regulatory. Finally, the most critical issues (and possible solutions) related to the management and orchestration of 5G with integrated satellite components were addressed in \cite{3gpp28808}. With respect to the NR 3GPP activities the SI ``Solutions for NR to support Non-Terrestrial Networks (NTN)" \cite{3gpp38821} was completed at the end of 2019. This SI completed the work that was initiated in Release 15 \cite{3gpp38811}. A set of required adaptations to enable NR technologies and operations in satellite networks were  addressed, covering several issues in RAN1 (Physical layer), RAN2 (Layer 2 and 3), and RAN3 (Interfaces).
    
    \item Release 17: Work on the SI entitled ``Study on architecture aspects for using satellite access in 5G" \cite{3gpp23737} continued and was last updated in March 2021. By the end of 2019, two WIs for NTNs had been initiated: (i) ``Solutions for NR to support NTN" \cite{3gpp38821}, under RAN activities; and (ii) ``Integration of satellite components in the 5G architecture" \cite{5gsatarch860005}, under SA. For the former, the activities are in the final stage and it was last updated on June 2021. However, the objective is the following points: (i) consolidation of the  impacts on the physical layer  and definition of potential solutions; (ii) evaluating the performance of NR in selected deployment scenarios (LEO based satellite access, GEO based satellite access) through link level (radio link) and system level (cell) simulations; and (iii) identification of the potential requirements for the upper layers based on the considered architectures. The latter WI goal was to extend the analysis provided in \cite{3gpp23737} through the following: (i) identification of impacted areas in NR systems due to  the integration of satellite components  in 5G; (ii) analysis of the issues related to the interaction between the core network and the RAN; and (iii) identification of solutions for the two highlighted use cases  (terrestrial and satellite network roaming and 5G fixed backhaul). Several areas were considered, such as network discovery and selection and network slicing. For Release 17, the freeze of the RAN1 physical layer specifications is scheduled to be in December 2021. This is to be followed by the Stage 3 freeze (RAN2, RAN3 and RAN4) by March 2022 and the ASN.1 freeze and the performance specifications completion is planned for September 2022 based on the timeline agreed back in December 2019.

\item Release 18: The June 28 – July 2, 2021 Workshop on Release 18, which is the start of 5G-Advanced, is to specify the topics of Release 18 with submissions divided into three areas (preliminary agenda): \acrfull{embb} driven work, Non-eMBB driven functionality and cross-functionality for both. This 3GPP workshop on the radio specific content of Release 18 reviewed over 500 presentations by companies and partner organizations, to identify topics for the immediate and longer-term commercial needs. The Release 18 Package Approval is expected  at the December 2021 Plenary (RAN\#94). The time duration for the release in RAN is tentatively set at 18 months. The detailed discussions on how to consolidate topics into WIs and SIs will not begin before RAN\#93-e meeting in September 2021. This meeting will see progress on ‘high-level descriptions’ of the objectives for each topic. The list of topics include the following \cite{rel8News}:
\begin{itemize}
   \item Evolution for  downlink \acrfull{mimo}
    \item Uplink enhancements
    \item Mobility enhancements
    \item Additional topological improvements (\acrfull{iab} and smart repeaters)
    \item Enhancements for \acrfull{xr}
    \item Sidelink enhancements (excluding positioning)
    \item RedCap evolution (excluding positioning)
    \item NTN evolution, including both NR and IoT aspects
    \item Evolution for broadcast and multicast services
    \item Expanded and improved positioning
    \item Evolution of duplex operation
    \item \acrfull{ai}/\acrfull{ml}
    \item Network energy savings
    \item Additional RAN1/2/3 candidate topics:
    \begin{itemize}
        \item Set 1: UE power savings, enhancing and extending the support beyond 52.6GHz, \acrfull{ca}/\acrfull{dc} enhancements (e.g., \acrfull{mrmc}, etc.), Flexible spectrum integration, \acrfull{ris}.
        \item Set 2: UAV, \acrfull{iiot}/\acrfull{urllc}, $<$5MHz in dedicated spectrum, other IoT enhancements and types,  HAPSs, Network coding.
        \item Set 3: Inter-gNB coordination, network slicing enhancements, multiple universal subscriber identity modules), UE aggregation, security enhancements, \acrfull{son}/\acrfull{mdt}.
    \end{itemize}
    \item Potential RAN4 enhancements.
\end{itemize}

\end{itemize}

\begin{figure*}[t]
\centering
\includegraphics[width=1\textwidth]{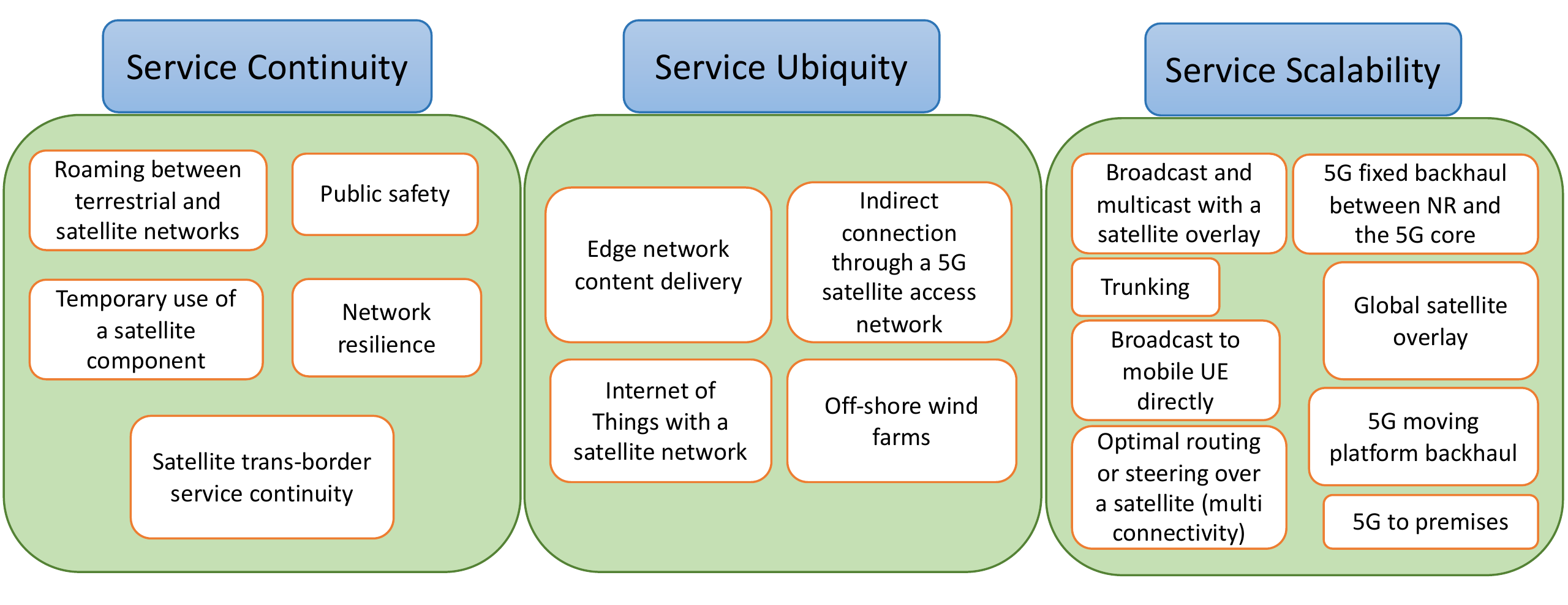}
\caption{Satellite access use cases in 5G. \textbf{Service continuity} use cases provide continuous access to services granted by the 5G system, while users move between terrestrial and satellite networks. Use cases considering fleets of such UE (whether locally grouped or dispersed) are also included in this category. \textbf{Service ubiquity} use cases serve potential users wishing to access 5G services in  ``unserved" or ``underserved" areas by terrestrial networks, which will be possible through 5G satellite access network service. \textbf{Service scalability} use cases utilize the distinguished capability of satellites in multicasting or broadcasting a similar content over a large area, and potentially directly to user equipment. Similarly, a satellite network can also contribute to off-loading traffic from terrestrial networks during  busy hours by multicasting or broadcasting non-time-sensitive data in non-busy hours.}
\label{UsecasesCategory}
\end{figure*}

The importance of the NTN evolution has been very visible in RAN meetings, including RAN\#93-e. In addition, satellite access might be discussed under other topics, such as mobility management and evolution for broadcast and multicast services. A summary of the 3GPP satellite related standardization activities is presented in Table \ref{ReleasesSummary}.

\begin{table*}[h]
\centering
\caption{Summary of 3GPP satellite related standardization activities.}
\label{ReleasesSummary}
\begin{scriptsize}
\begin{tabular}{|p{1.5cm}|p{1.5cm}|p{1.5cm}|p{5cm}|p{4cm}|}
\hline
\rowcolor[HTML]{CBCEFB} 
Release & Item code & TSG & Title & Status \\ \hline
\cellcolor[HTML]{DAE8FC}Release 15 & TR 38.811 & RAN & Study on NR to support NTN & \begin{tabular}[c]{@{}l@{}}Completed \\ October 2020\end{tabular} \\ \hline
\cellcolor[HTML]{DAE8FC}Release 16 & TR 22.822 & SA & Study on using satellite access in 5G & \begin{tabular}[c]{@{}l@{}}Completed \\ June 2018\end{tabular} \\ \hline
\multicolumn{1}{|c|}{\cellcolor[HTML]{DAE8FC}} & \begin{tabular}[c]{@{}l@{}}TR 23.737\\ First stage\end{tabular} & SA & Study on architecture aspects for using satellite access in 5G & \begin{tabular}[c]{@{}l@{}}Started Release 16, June 2018\\ Last update Release 17, March 2021\end{tabular} \\ \cline{2-5} 
\multicolumn{1}{|c|}{\cellcolor[HTML]{DAE8FC}} & \begin{tabular}[c]{@{}l@{}}TR 23.737\\ Second stage\end{tabular} & SA & Integration of satellite components in the 5G architecture & \begin{tabular}[c]{@{}l@{}}Started June 2020 \\ Last update June 2021\end{tabular} \\ \cline{2-5} 
\multicolumn{1}{|c|}{\cellcolor[HTML]{DAE8FC}} & TR 28.808 & SA & Study on management and orchestration aspects with integrated satellite components in a 5G network & Completed March 2021 \\ \cline{2-5} 
\multicolumn{1}{|c|}{\multirow{-4}{*}{\cellcolor[HTML]{DAE8FC}Release 17}} & TR 38.821 & RAN & Solutions for NR to support NTN & \begin{tabular}[c]{@{}l@{}}Started Release 16, June 2018\\ Last update Release 17, June 2021\end{tabular} \\ \hline
\cellcolor[HTML]{DAE8FC} & 920035 (5GSATB) & SA & 5G system with satellite backhaul & Completed June 2021 \\ \cline{2-5} 
\multirow{-2}{*}{\cellcolor[HTML]{DAE8FC}Release 18} & 920034 (SCVS) & SA & 5G system with satellite access to support control and/or video surveillance & Started June 2021 \\ \hline
\end{tabular}
\end{scriptsize}
\end{table*}


\subsection{Use cases for satellite access in 5G} \label{sec32}

In March 2017, as part of Release 14, the 3GPP initiated an SI to analyze the feasibility of satellite integration
into 5G network (3GPP TR 22.822 \cite{3gpp22822}). The initial goal was to bring together satellite operators and other companies to create aligned contributions
in the support of satellites in the 5G standardization. The two SIs 3GPP TR 38.811 and 3GPP TR 38.821 had been already completed. They studied the role of satellites in the 5G ecosystem. In addition, the challenges of a co-existing satellite-terrestrial network had been analyzed taking into account different architectural  options. 


On the basis of satellite networks characteristics introduced in Section \ref{sec2} above, three main use case categories are defined, namely service continuity, service ubiquity, and service scalability, as shown in Figure \ref{UsecasesCategory}. Table \ref{Table2} summarizes the satellite access use cases in \cite{3gpp38811} and \cite{3gpp22822} and their applications. To realize these use cases, some requirements need to be fulfilled by 5G+ systems:
\begin{itemize}
    \item The 5G system shall support service continuity between land-based 5G access and satellite-based access networks owned by the same operator or by an agreement between operators with guaranteed QoS while switching to or from terrestrial to satellites.
    \item Provide the optimum network selection. 
    \item Support mMTC and Nb-IoT services.
    \item The 5G system shall support the use of satellite links between the radio access network and core network and within the core network by enhancing the 3GPP system to handle the latencies introduced by satellite backhaul.
    \item Reciprocal cooperation is required between mobile operators and satellite operators to ensure good service areas available for customers. 
    \item  A 5G system supporting 5G satellite access and 5G terrestrial access shall be able to optimally distribute user traffic over both types of access.
    \item A 5G system providing service with satellite access shall be able to support QoS indicators adapted to \textcolor{black}{GEO-based satellite access with RTT of  600-800 ms, MEO-based satellite access with RTT of 125-250 ms, and LEO-based satellite access with RTT of 30-50 ms \cite{telesat}. These RTT values include the delays of processing on both ground and orbit as well as the variable propagation delays. In satellite communication, propagation delay varies due to changes in satellite and user positions, which lead to different slant ranges.}
    \item UE with satellite access shall have the capability to accept or reject connections with the satellite on the basis of the quality of class of indicators that are supported, and on the basis of the available accesses.
    \item A 5G system with multiple access shall be able to select the combination of access technologies to serve UE on the basis of the targeted priority, pre-emption, QoS parameters and access technology availability.
\end{itemize}

\begin{footnotesize}
\begin{table*}[h!]
\caption[Satellite access use cases in 5G]{Satellite access use cases in 5G.}\label{Table2}
\centering
\begin{tabular}{|p{3.3cm}|p{8cm}|p{4cm}|}
\hline 
\rowcolor[HTML]{CBCEFB} \textbf{Use case} & \textbf{Description} & \textbf{Application}\\
\hline 
Roaming between terrestrial and satellite networks \cite{3gpp22822}& UE will have worldwide connectivity via terrestrial and/or satellite networks while roaming between different terrestrial and satellite operators. Optimal network selection will be possible when both terrestrial and satellite networks are available. & Tracking shipping containers worldwide.\\
\hline
Broadcast and multicast with a satellite overlay \cite{3gpp22822} & 
The high demand for digital content distribution services by a large number of UEs might lead to the saturation of the transmission capabilities of the mobile network operator (MNO). A satellite network operator (SNO) can boost the capacity of terrestrial MNOs and serve UEs that are located beyond terrestrial network coverage. & TV broadcasting and video streaming. \\
\hline
Edge network content delivery \cite{3gpp38811} &	Offloading popular  media and entertainment content from the mobile network infrastructure and delivering it at the network edge, where it may be stored in a local cache or further distributed to the UEs.
& Broadcast channel to support multicast delivery to 5G network edges.\\
\hline
Broadcast to mobile UE directly \cite{3gpp22822}& Public safety authorities want to be able to instantaneously alert the public of catastrophic events and provide guidance during disaster relief. Also media and entertainment industry can provide entertainment services in vehicles (cars, buses, trucks).	& Broadcast or multicast service directly to UE whether handheld or vehicle mounted. \\
\hline
Public safety \cite{3gpp38811} &	To achieve continuity of services for  emergency responders, such as police, fire brigade, and medical personnel, while exchanging messages and voice services in outdoor conditions anywhere and  under any mobility scenarios.	& Access to emergency responder equipment (handset or vehicle mounted). \\
\hline
Internet of Things with a satellite network \cite{3gpp22822}& An IoT service provider uses a constellation of LEO satellites that can provide continuous global coverage for UEs with limited RF and energy capabilities. &  Reporting items or people positions on a continuous basis for security and tracking purposes. \\
\hline
Temporary use of a satellite component \cite{3gpp22822}& When a disaster occurs, elements of the 5G RATs may be partially or completely destroyed. The normal terrestrial network access may no longer be available..
However, communication services are required to provide first aid, emergency support, restore security, and organize logistics. 5G satellite RATs can be used to provide the required access to 5G networks. & Significant earthquake, flood, or war.\\
\hline
Network resilience \cite{3gpp38811} & To prevent a complete network connection outage on critical network links that require high availability. & Secondary or backup connection (although potentially limited in capability compared to the primary network connection).\\
\hline
Trunking \cite{3gpp38811}& A network operator may want to interconnect various 5G local access network islands. & Industrial sites. \\
\hline

Optimal routing or steering over a satellite (multi-connectivity) \cite{3gpp22822}& At the edge of the radio coverage of a 5G terrestrial RAT, the performance of eMBB and mMTC services might be limited at certain times. Delay-insensitive communication can be routed through satellites whereas delay-sensitive communication can be achieved through terrestrial 5G RATs. & Automated factories or industrial sites in remote areas.\\
\hline
Satellite trans-border service continuity \cite{3gpp22822}& 5G terrestrial coverage is not always available near countries borders, and UEs crossing borders need to switch from one operator to another. A 5G satellite access network covering border areas can provide continuous coverage and support a smooth transition from one operator to another.  & Communication coverage for travellers and intelligent transportation systems.\\
\hline
Global satellite overlay \cite{3gpp22822}& When the distance between two sites increases, the difference in latency between air and optical fibre transmission media may become critical for some applications. A constellation of LEO satellites, where each satellite is equipped with a gNB and interconnected with other neighbouring satellites via ISLs, provides an overlay mesh network for users that have a need for long distance connectivity with improved latency performance or specific end-to-end security. & Critical application domains, such as High Frequency Trading (HFT), Banking or Corporate communications for global organizations with distributed sites around the world.  \\
\hline
Indirect connection through a 5G satellite access network \cite{3gpp22822}& UEs with no direct access to the 5G network or satellite communication can access a 5G network through a relay enabled UE and a bent-pipe satellite-enabled or 5G satellite-enabled interconnection.  & UEs on a commercial jet or maritime cruise vessel.\\
\hline
5G fixed backhaul between NR and the 5G Core \cite{3gpp22822}& Satellite backhaul provides a cost efficient option for mobile operators to provide services in rural areas where terrestrial infrastructure is not available. The site of a cell tower in a rural area can be connected to the 5G core through satellite backhaul.  & Remote and isolated villages with low-density populations. \\
\hline
5G moving platform backhaul \cite{3gpp22822}& To provide a backhaul connection for 5G base stations mounted on trains or airplanes where there is no 5G terrestrial coverage. & Internet and communication services for users on-board trains or planes. \\
\hline
5G to premises \cite{3gpp22822}& A terrestrial cellular operator  works with a satellite operator to provide better services for customers in unfavourable geographical areas with old terrestrial network infrastructure, using a new home or office gateway unit to combine the available signals from satellite and terrestrial networks, and to present good WIFI coverage within the premises. Satellites are used to broadcast and multicast media. Caching can be done on the gateway. Unicast will use the cellular route, especially for delay-sensitive applications & Supports Internet and multimedia intensive applications in rural and remote areas. \\
\hline
Off-shore wind farms \cite{3gpp22822}& At an off-shore wind farm, the wind power plant communication network  connects to the on-shore and inland remote
service centre through a 5G satellite connection. & Connecting remote off-shore wind farms to the service centre.  \\
\hline
\end{tabular}
\end{table*}
\end{footnotesize}

\section{Satellite Access in 5G and Its Architecture Aspects} \label{sec4}

In TR 38.811 \cite{3gpp38811}, the 3GPP community introduced two types of satellite access networks: 
\begin{itemize}
    \item A broadband access network serving \acrfull{vsat} that can be mounted on a moving platform (e.g. bus, train, vessel, aircraft). In this context, broadband refers to at least 50 Mbps data rate and even up to several hundreds Mbps for downlink. The service links operate in frequency bands allocated to satellite and aerial services (fixed, mobile) above 6 GHz.
    
    \item Narrow- or wide-band access network serving terminals equipped with omni- or semi-directional antenna (e.g., handheld terminal). In this context, narrow-band  refers to less than a 1 or 2 Mbps data for downlink. The service links operate typically in frequency bands allocated to mobile satellite services below 6 GHz.
\end{itemize}

\begin{figure}[h]
\centering
\includegraphics[width=0.5\textwidth]{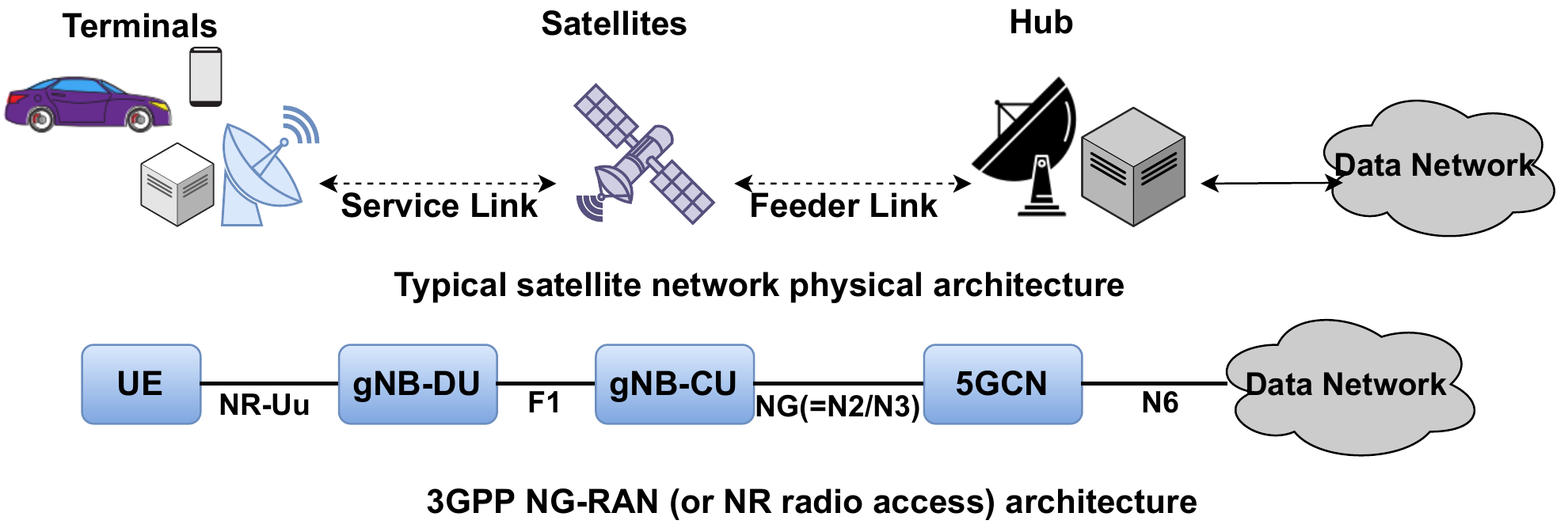}
\caption{Comparison between the NG-RAN logical architecture and the satellite network physical architecture.}
\label{ComparisonArchit}
\end{figure}

In terms of architecture, Figure \ref{ComparisonArchit} shows a comparison between a typical satellite network physical architecture and a 3GPP NG-RAN architecture. To integrate satellite access networks in 5G, 3GPP TR 38.821 introduced the following three types of satellite-based NG-RAN architectures:

\begin{itemize}
    \item \textbf{Transparent satellite-based NG-RAN architecture:} The satellite payload implements frequency conversion and a radio frequency amplifier in both uplink and downlink directions. Several transparent satellites may be connected to the same gNB on the ground.
\item \textbf{Regenerative satellite-based NG-RAN architectures:} The satellite payload implements regeneration of the signals received from earth. The satellite payload also provides \acrshort{isl}s between satellites.  An ISL may be a radio interface or an optical interface that may be 3GPP or non-3GPP defined. The regenerative satellite-based NG-RAN architecture has two types:
\begin{itemize}
    \item gNB processed payload
    \item gNB-DU processed payload
\end{itemize}

\item \textbf{Multi-connectivity involving satellite-based NG-RAN:} This may apply to transparent satellites as well as regenerative satellites with gNB or gNB-DU function on board.
\end{itemize}

\begin{figure}[h]
\centering
\includegraphics[width=0.5\textwidth]{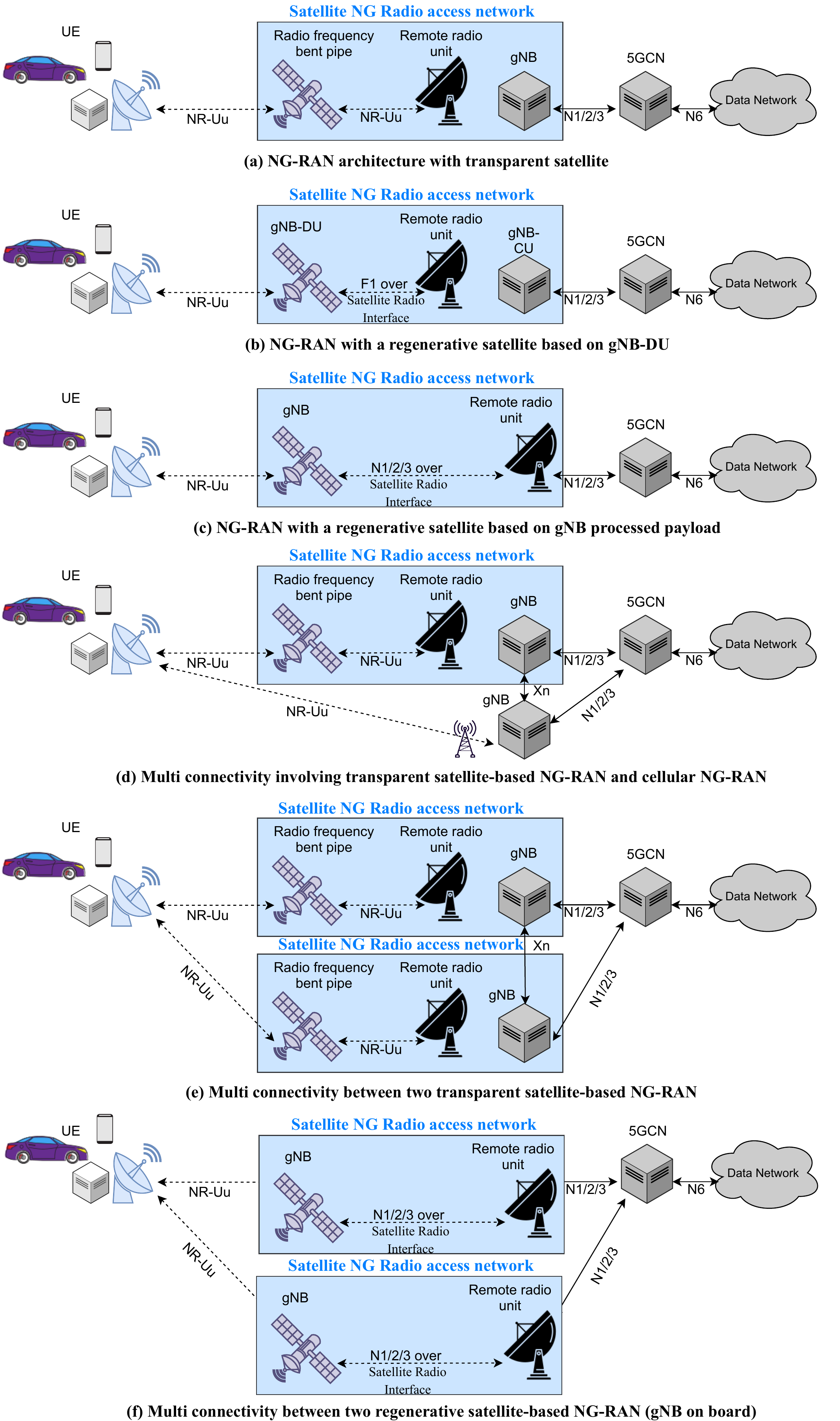}
\caption{The three types of satellite-based NG-RAN architectures described in \cite{3gpp38821}. }
\label{MappingOption}
\end{figure}

Figure \ref{MappingOption} illustrates the aforementioned  satellite access network architectures and how the satellite access components are mapped onto the 5G architecture.

\section{Adapting New Radio for Satellite Networks} \label{sec5}

The technical reports 3GPP TR 38.811 \cite{3gpp38811} and 3GPP TR 38.821 \cite{3gpp38821} discussed channel modeling for satellites, where channel model parameters were provided while taking different user environments and atmospheric conditions into consideration. Some design constraints are identified and the impacted areas of NR are explained in the following two subsections.

\subsection{Specific constraints associated with satellite networks}\label{sec51}
Compared to cellular systems, there are some design constraints that need to be addressed when considering satellite network deployment scenarios:
 
\begin{itemize}
    \item \textbf{Propagation channel:} the channel has a different multi-path delay and Doppler spectrum model. However, for narrowband signals and frequency bands below 6 GHz, the time disparity may be ignored. Certain outdoor conditions and line-of-sight operations are necessary for the UE communication via satellite. 
    \item \textbf{Frequency plan and channel bandwidth:} The allocated spectrum to a satellite system is respectively 2 x 15 MHz (\acrfull{ul} \& \acrfull{dl}) at S band and about \textcolor{black}{2 x 2.5 GHz for UL and 2 x 2.4 GHz for DL} at Ka band. In addition, satellite systems at S and Ka bands use mostly circular polarizations. With frequency re-use and efficient spectrum allocation among different cells, the maximum channel bandwidth per cell is assumed to be respectively 2 x 15 MHz (UL and DL) at S band and \textcolor{black}{up to 2 x  2.4 GHz} (UL and DL) at Ka band. However, inter cell interference should be minimized.
    \item \textbf{Power limited link budget:} Two main design aspects need to be considered:
    \begin{itemize}
        \item Maximizing the throughput for a given transmit power from the UE on the UL and from the satellite on the DL.
        \item Maximizing the availability of the service under deep fading situations (typically between 20 and 30 dB in Ka band for 99.95\% availability).
    \end{itemize}
    \item \textbf{Cell pattern generation:} Satellites have larger cells compared to cellular networks, and the cells are moving in case of NGSO satellite. This creates a significant difference in propagation delay between UE at the cell edge and UE at the cell centre, and the difference in propagation delay increases as the altitude of the satellite decreases. Accordingly, when the position of UE is not known by the network, contention-based channel access might be impacted.
    
    \item \textbf{Propagation delay characteristics:} Satellite systems have much greater propagation delays than terrestrial systems, which can impact all round-trip signaling times, especially at access and transport (data transfer) levels.
    
    \item \textbf{Mobility of the infrastructure's transmission equipment:} For GSO satellites, the transmission equipment is quasi static with respect to the UE with only small Doppler effects. For NGSO satellites, by contrast, move relative to the earth and create higher Doppler effects than GSO systems. The Doppler depends on the frequency band and the relative satellite velocity with respect to the UE.  The Doppler effect will continuously modify the carrier frequency, phase, and spacing. However, most of the Doppler shift and variation rate can be compensated by utilizing the predictable motion of  satellites as well as UE location if known.
    
    \item \textbf{Service continuity between land-based 5G access and non-terrestrial-based access networks:}  Different handover triggering mechanisms are required in order to give preference to  cellular communication. The handover procedure should support both regenerative  and bent-pipe satellites. In addition, it should handle handover preparation and failure while also supporting lossless handover. Handover can be due to intra-non-terrestrial network mobility as well as between non-terrestrial and cellular networks.

\item \textbf{Radio resource management adapted to network topology:} Unlike cellular systems where access control is typically located close to the UE, in satellite systems access control is mostly located at the satellite base station, gateway, or hub level, which may prevent an optimal response time for access control. Hence, pre-grants, \acrfull{sps}, and/or a grant-free access scheme would be beneficial.

 \item \textbf{Terminal mobility:}  Very high speed UE with speeds of up to 1,000 km/h need to be supported.
\end{itemize}

\begin{table*}[t]
\caption{Impacted NR areas and potential solutions to support satellite networks.}
\label{Table1}
\begin{scriptsize}
\begin{tabular}{|p{1cm}|p{1cm}|p{1.5cm}|p{8cm}|p{5cm}|}
\hline
\rowcolor[HTML]{CBCEFB} Satellite network specifics & Effects & Impacted NR features & Potential adaptations & Potential areas of impact to be further studied \\
\hline
\multirow{4}{=}{Motion of the space vehicles (especially for Non-GEO-based access network)} & Moving cell pattern & Handover paging  & The NR UE may need to be capable of reporting its geo-location to the satellite RAN. The ephemeris information of the NGEO satellites can be used to determine the satellite or beam footprint. Thus, the network can determine which satellite/beam covers that location of a  UE  at any given time and for which duration. & Taking advantage of the knowledge of the UE location and satellite ephemeris information to adapt the NR handover and paging protocols needs further study. The NR beam management might assume the same frequency on the adjacent beams, but on the same satellite adjacent beams  may use different frequencies or different polarization, which would require the NR beam management procedures to be modified. \\
\cline{2-5}
& Delay variation  & Tracking area (TA) adjustment & Knowing the satellite orbits and UE position, the delay variation is quite predictable. Also, the transmission timing of the UE has to be adjusted over the borders of individual TTIs. & Solutions need to be studied to ensure the alignment of uplink signals over the satellite links to overcome the predictable delay of SatNets.   \\
\cline{2-5}
& \multirow{2}{=}{Doppler} & Synchronization in downlink & To meet the condition of 5ppm, the satellite altitude has to be above 13,000 km. For the Doppler shift amplitude to be compensated, it has to be less than 48 kHz for S band and 480 kHz for Ka band  & In case the aforementioned conditions are not met, further studies are required to accommodate the high Doppler shift during the cell synchronization procedure in SatNets. \\
\hline

\multirow{3}{=}{Altitude} &
\multirow{3}{=}{Long latency} & Hybrid automatic repeat request (HARQ) & Dealing with a higher number of parallel HARQ processes and its feasibility for non-terrestrial networks have been addressed in \cite{Kodheli2017} \cite{3gpp38821}. & The impacts of SatNet delays on the existing HARQ-supporting mechanisms need to be studied. In particular, what needs to be examined is the impact of long RTT delays on NR HARQ operations. Also the impact on UEs and serving satellite gNBs should be considered when the number of HARQ processes is either extended to satisfy high reliability scenarios or limited/disabled for longer SatNet delays. \\
\cline{3-5}
& & MAC/Radio Link Control (RLC) Procedures  & NR UE and base stations must size their transmission buffers and the retransmission time-out mechanisms according to the longest RTT to be anticipated. The number of retransmissions allowed before a packet is dropped from the retransmission buffer may also be adjusted.
UL scheduling delay parameters are expected to be redefined to accommodate the RTT of the associated deployment scenario.
 & No further study required.\\
\cline{3-5}  & & Physical layer Procedures (ACM, power control)  & For LEO satellites, ACM may also be used to adapt to the large variation of free space loss.  The variation is sufficiently slow compared to the 20 ms worst case RTT.  ACM should also be able to react to shadowing fades to a large extent, but still unable to follow fast fading. & Further studies are needed to define the required margin for power and ACM control loops to cope with the long RTT. \\
\hline
Cell size & Differential delay & Time advance in random access response (RAR) message  & For satellites, it is expected that a time advance mechanism will need to be modified to compensate for the propagation delay. & Further study is required as time advance should compensate for differential delay/distance.  \\

\cline{3-5} & & Physical random access channel (PRACH) & The current window for the PRACH response in NR cannot cover the long RTT. Therefore, the random access response window length in NR should be revisited to accommodate the RTT of satellites. However, extending the window size in the existing procedure introduces unnecessary UE monitoring intervals. For satellites, the current NR preamble format needs to be extended considering different footprint of satellite cells.  & The solution to handling long RTT with the consideration of power saving at the UE side needs to be further studied. Further studies should be done if a new random access preamble format is needed for satellites. \\
\hline
Duplex mode  & Regulatory constraints & Access scheme (time division duplex (TDD)/ frequency division duplex (FDD))  & When considering satellite networks, this guard time should equal the round trip delay. Although this guard time may be acceptable in the case of LEO access system, there is a need to deal with the variable delay. & The applicability of the duplexing mode TDD or FDD depends on the regulations (ITU-R and/or national) associated with the targeted spectrum. In case the regulations allow it, TDD mode can be considered for LEO-satellite-based access network with some required NR modifications. \\
\hline
\multirow{2}{=}{Satellite Payload performance} & Phase noise impairment & Phase tracking reference signal (PT-RS) & PTRS is needed in NR supporting SatNets for phase error compensation. Typical phase noise masks of state-of-the-art bent pipe of satellite payloads in non-terrestrial networks can be efficiently compensated by the current NR design in the absence of important Doppler shifts and/or residual CFO at a carrier frequency of up to 30 GHz. It has been proposed that eMBB should be supported by SatNets and the modulation order might not always be low. PT-RS is needed in SatNets at high carrier frequencies, where the maximum received SNR is limited by the phase noise. & Considering the high speed of  LEO satellites, some further investigation is needed in the case of important Doppler shifts and/or residual CFO, or in the presence of the specific phase noise masks of on-board payloads that are significantly different from the ones considered in cellular networks so far, or with very large channel bandwidths. \\
\cline{2-5}
 & Back-off  & Peak-to-average power ratio (PAPR)  & The use of CP-OFDM for downlink does not restrict NR operation in SatNets, but it may affect the system performance. When a satellite transponder is sufficiently wide and powerful to accommodate more than one FDM carrier, each of which is a different NR CP-OFDM signal, the satellite amplifier is backed off to minimize the intermodulation between these FDM carriers within the same transponder.  As a result, the distortion introduced is small. For communicating with small UEs, the satellite amplifier is used to send only one NR CP-OFDM downlink.  It is highly desirable to operate the amplifier with as small an output power backoff (OBO) as possible.  But, due to the higher PAPR of CP-OFDM signal, sufficient OBO is necessary.  To close the link, it may be necessary to reduce the size of CP-OFDM carrier or to operate the CP-OFDM carrier with a lower modulation and coding mode.  Either way, the forward link capacity is reduced significantly. On the uplink, DFT-spread-OFDM might be beneficial.

  & PAPR reduction techniques of CP-OFDM signal for downlink would be beneficial for optimizing the capacity of non-terrestrial networks and therefore could be considered in future studies. \\
 \hline
Network architecture & RAN Mapping & Protocols  & In \cite{3gpp38811} three mapping options are described that prevent the need to create new interfaces or reference points, as described in Figure \ref{MappingOption}. The timers associated to the protocols transported over the feeder and service links (e.g., F1, N1, N2 and N3 reference points) may need to be extended. 
In the processed payload option, it is recommended that the satellite implements some kind of ``intermediate" node, which could be based on the outcomes of  3GPP TR 38.874 \cite{3gpp38874}, and see how it can be applicable and what areas of impact it may create. Location update, paging, and handover RAN related protocols need to be adapted depending on tracking area design. Also, handling of network identities needs to be adapted. &  The N1/N2/N3 and GTP-based F1 interface protocols may need to be adapted to accommodate the satellite feeder link characteristics (long delay, BER). Depending on the SatNets deployment scenarios, other impacts on NR specification may have to be considered, such as location update, paging, and handover RAN related protocols and handling of network identifies.\\
\hline
\end{tabular}
\end{scriptsize}
\end{table*}

\subsection{NR impacted areas and potential solutions}\label{sec52}

The aforementioned satellite related design constraints have impacts on some features of NR as documented in 3GPP TR 38.811 \cite{3gpp38811} and 3GPP TR 38.821 \cite{3gpp38821}. The following points provide a brief description of how NR is impacted, and 
Table \ref{Table1} summarizes the potential solutions to support NR in satellite networks .

\begin{itemize}
\item \textbf{Handover paging:} The fast movement of LEO satellites and their many beams mean that UEs are only kept within a beam for a few minutes. The rapid change creates problems for paging as well as for handovers for both stationary UEs and moving UEs. A handover has to be executed quickly otherwise the UE may not make use of the satellite resources efficiently and may suffer a loss of data. With fixed tracking areas on the ground, there is no one-to-one correspondence between moving beams and fixed tracking areas or registration areas, which is necessary for the paging process.

\item \textbf{\acrfull{ta} adjustment:} Moving satellites generate strong delay variations, and a fast change in the overall distance of the radio link between the UE and BS via satellite. This delay largely exceeds the \acrfull{tti} of NR, which is equal to or less than 1 ms. Hence, the TA alignment is an important feature of NR that will be impacted by the introduction of satellites in 5G to ensure that all uplink transmissions are synchronized at gNB reception point.

\item \textbf{Synchronization in downlink:} In order to access the 5G network, the UE has to detect the primary  and secondary synchronization signals. These synchronization signals allow time and frequency correction as well as Cell ID detection. UE in cellular network has to get good one-shot detection probability at -6 dB received baseband \acrfull{snr} condition with less than 1\% false alarm rate, with robustness against initial frequency offset up to 5 ppm. It is expected that these requirements defined for terrestrial UE will be kept the same for SatNets UE. Even though   the SNR level of satellite systems is typically in the range of -3 to 13 dB SNR, the satellite movement creates a higher Doppler shift, depending on the frequency band and the velocity of the satellite relative to the UE. \textcolor{black}{However, this can be compensated for at the demodulator.} 

\item \textbf{\acrfull{harq}:} The HARQ process is a time-critical mechanism. In SatNets, the \acrfull{rtt} normally exceeds the maximum conventional HARQ timers or the maximum possible number of parallel HARQ processes. This means that simply extending the number of HARQ processes linearly to RTT might not be feasible for some UE due to memory restrictions and the maximum possible parallel processing channels. Also, the impact of this delay has to be considered by the gNBs on the number of their active HARQ processes. Although NR has extended the number of HARQ processes in Rel. 15 to 16 processes, for SatNet NR the number of HARQ processes may need to be further extended flexibly according to the induced RTT delay.

\item \textbf{MAC/\acrfull{rlc} procedure:} For LEO satellite systems, the one way propagation delay changes continuously (e.g., 2-7 ms for 600 km orbit). The \acrfull{arq} requires that the transmitted packets be buffered and released only after the successful receipt of an acknowledgement or until a time out. A larger transmission buffer is required due to the long RTT, which also limits the number of retransmissions allowed for each transmitted packet. The ARQ transmit buffer size and retransmission mechanism must be designed for the longest possible delay (i.e., at the lowest elevation). Scheduling mechanisms must be able to cope with the long RTT.

\item \textbf{Physical layer procedure (\acrfull{acm}, power control):} Due to the large free-space loss and limited power available at the UE and satellites, the power margin is limited.  Thus, only a limited amount of power control is available for satellite links. Due to the long delay in the loop, the power control is not expected to track fast fading, but may be used to track slower power variations. The slow reaction time, due to long RTT, is expected to impact the performance of some physical layer procedures, particularly those with close control loops, such as power control and ACM. However, most control loops require some adjustments in implementation, but not a fundamentally different design.

\item \textbf{Time advance in \acrfull{rar} message:} Time advance mechanisms ensure that transmissions from all UE operating in the same cell are synchronized when received by the same gNB. A time advance command is provided to the UE in an RAR message during initial access and later to adjust the uplink transmission timing. The maximum value of the time advance command constrains the maximum distance between UEs and base station, which defines the allowed cell size. 

\item \textbf{\acrfull{prach}:} It is necessary to consider the long RTT impact on PRACH. For a given beam covering a cell, there is one common propagation delay for all served UE, and  one relative propagation delay for each served UE. If the common propagation delay can be compensated, then the satellite PRACH signal design will depend on the relative propagation delay, which is limited to a TA range of up to 200 km in current NR specifications. However, when the TA is of thousands of kilometers a satellite PRACH signal and procedure design need to be modified.

\item \textbf{Access scheme (\acrfull{tdd}/ \acrfull{fdd}):} Most existing satellite systems operate in the frequency bands designated for the FDD mode with a defined transmit direction.  For some frequency bands, the TDD mode is possible. 
When considering the TDD mode, a guard time is necessary to prevent the UE from simultaneously transmitting and receiving. This guard time directly depends on the propagation delay between the UE and gNB. This guard time will directly impact the useful throughput and hence the spectral efficiency.

\end{itemize}

\begin{figure*}[h]
\centering
\includegraphics[width=1\textwidth]{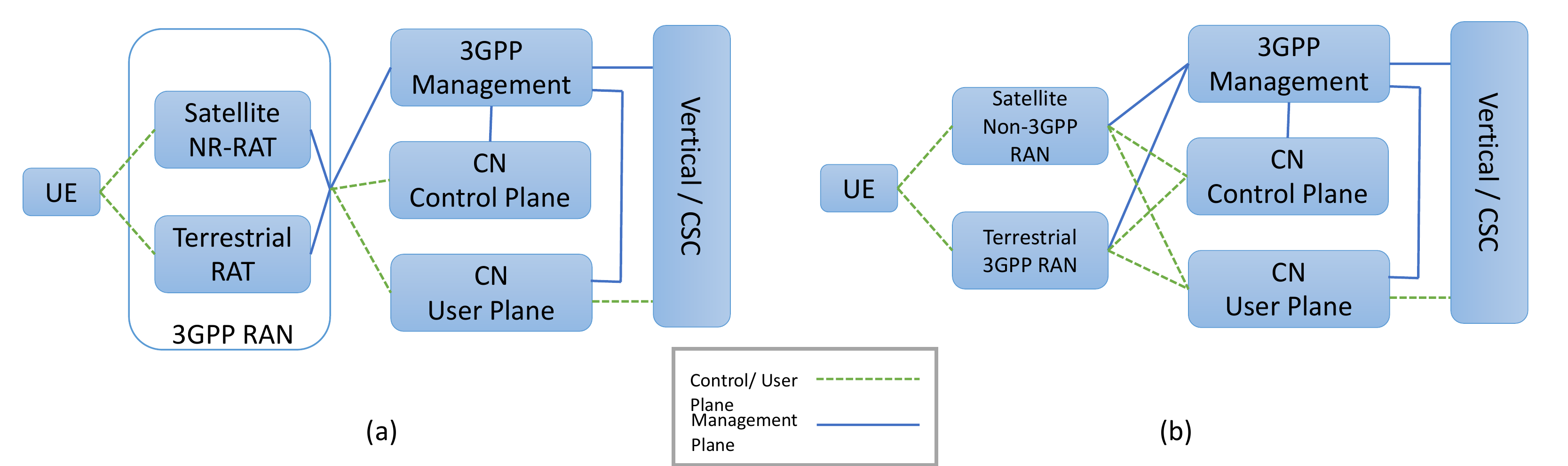}
\caption{Reference management architectures for integrated satellite components as described in \cite{3gpp28808}. (a) Reference architecture for the management of a 3GPP RAN integrating a satellite NR-RAT with a terrestrial RAT. (b) Reference architecture for the management of a non-3GPP satellite RAN integrated in a 5G network.}
\label{ManagementArch}
\end{figure*}

\begin{itemize}

\item \textbf{Phase tracking reference signal (PT-RS):} Phase variations in time domain can be caused by different phenomena, including the presence of phase noise, frequency drifts due to Doppler shift, or due to insufficient frequency synchronization (e.g., residual CFO), etc. In NR, \acrshort{ptrs} has been introduced to compensate for phase errors. The PT-RS configuration in NR is very flexible and allows user-specific configurations depending on scheduled MCS/bandwidth, UE RF characteristics, demodulation reference signal configuration, waveform, etc. PT-RS configuration flexibility is beneficial for SatNets.

\item \textbf{Peak-to-average power ratio (PAPR):} A key component in satellite payload architecture is a power amplifier. It exhibits nonlinear behavior when operating near saturation in an effort to increase power efficiency. Nonlinear distortion causes constellation warping and clustering, thus complicating signal reception. \acrshort{papr} is a measurement that determines the vulnerability of the transmitted signal to nonlinear distortion, where higher values indicate a worse impact. In the NR downlink, \acrfull{cpofdm} is used resulting in higher PAPR values compared with the underlying modulation in a single carrier. The distortion can be reduced by increasing the backoff of the amplifier operating point. But this reduces the amplifier efficiency accordingly. \textcolor{black}{For satellite communications, low PAPR waveforms are desired. }

\item \textbf{Protocols:} Mapping is needed between the NG-RAN logical architecture and the SatNet architecture (the two architectures are depicted in Fig. \ref{ComparisonArchit}). Several mobility scenarios should be considered, specifically the mobility induced by the motion of satellites, the motion of UEs from one  beam to another beam  generated by the same satellite, the motion of UEs between beams generated by different satellites, and the motion of UEs between satellites and cellular access. Location updating, paging, and handover RAN related protocols need to accommodate the extended delay of intra-satellite access mobility, the differential delay when mobility is between a satellite and a cellular network, and the mobility of the cell pattern generated by NG-satellites.

\end{itemize}

\section{Management and Orchestration with Integrated Satellite Components in a 5G Network}\label{sec6}

Orchestration allows network services and resources to be managed and controlled on an integrated basis and optimized. Thus, time spent on processes is shortened by rapidly and flexibly allocating network components and resources. In 2019, the 3GPP TSG SA started a study on management and orchestration aspects with integrated satellite components in a 5G network. The main objective was to study business roles as well as service, network management, and orchestration of a 5G network with integrated satellite components.  The scope of the study covered both NTN RAN-based satellite access, non-3GPP defined satellite access, as well as backhaul aspects. The study outcome was presented in TR 28.808 \cite{3gpp28808} which included  potential requirements and solutions to integrate satellites in 5G networks, such as network slice management and management and monitoring of gNB components.

\begin{figure*}[h]
\centering
\includegraphics[width=1\textwidth]{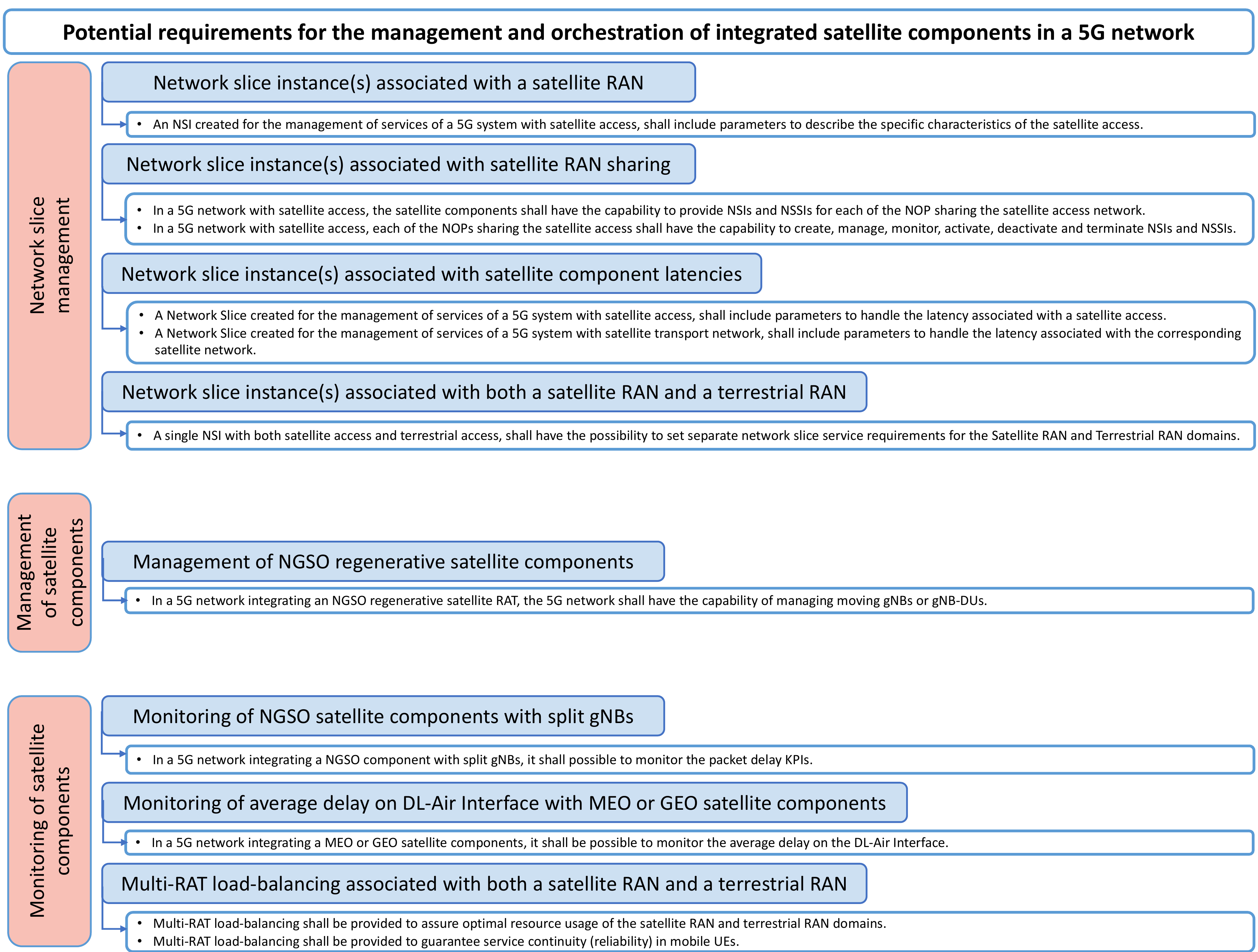}
\caption{Potential requirements for the management and orchestration of integrated satellite components in a 5G network.}
\label{ReqManagement}
\end{figure*}

The study presented two reference management architectures for integrated satellite components, as shown in Figure \ref{ManagementArch}. The first reference architecture was for the management of a 3GPP RAN integrating a satellite NR-RAT with a terrestrial RAT. The second reference architecture was for the management of a non-3GPP satellite RAN integrated in a 5G network. The potential requirements for the management and orchestration of integrated satellite components in a 5G network are shown in Figure \ref{ReqManagement}. The requirements are presented in three categories: (1) network slice management requirements; (2) management of satellite components; and (3) monitoring of satellite components. 

Compared to terrestrial NR, the impact of integrating satellites mainly comes from LEO/MEO scenarios where gNB components, such as gNB-DU, are located onboard satellites and therefore move faster than the earth. Other enhancements are required due to the long delays and RTT that impact some of the monitoring functionality and key performance indicators of 5G networks. The study concluded that the concepts of Self-Organizing Networks (SONs) for 5G would need to be enhanced to support mobile non-terrestrial gNBs. Although efficient network management is essential in future integrated networks in order to fully utilize the available network resources, the standardization work on NTN management is nevertheless quite limited within the 3GPP working groups.


\section{Other Standardization Organizations} \label{sec7}

\paragraph{\textbf{\acrfull{itu} Standardization}}
In ITU terminology, the 5G system is called IMT-2020. The IMT-2020 network architecture is envisioned to be access network-independent and with a core network common to \acrfull{rat} for IMT-2020, as well as existing fixed and wireless networks. The IMT-2020 core network control mechanisms will be decoupled from the access network technologies. The IMT-2020 network should support new RATs for IMT-2020, evolved IMT-advanced RATs, satellite networks, fixed broadband network access, and \acrfull{wlan} access networks. In July 2020, 3GPP 5G has been formally endorsed as ITU IMT-2020 5G standard. The M family of ITU-R recommendations refers to 5G systems.  A series of ITU-R recommendations (standards) for the satellite components for IMT have been already developed, which includes the integration of terrestrial and satellite mobile communication systems \cite{ITUM}:
\begin{itemize}
    \item Rec. ITU-R M.1167 - ``Framework for the Satellite Component of International Mobile Telecommunications-2000 (IMT-2000)".
    \item Rec. ITU-R M.818 - ``Satellite Operation within International Mobile Telecommunications-2000 (IMT2000)".
    \item Rec. ITU-R M.1850 - ``Detailed Specifications of the Radio Interfaces for the Satellite Component of International Mobile Telecommunications-2000 (IMT-2000)".
    \item Rec. ITU-R M.1182 - ``Integration of Terrestrial and Satellite Mobile Communication Systems".
    \item Rec. ITU-R M.2014 - ``Global Circulation of IMT-2000 Satellite Terminals".
\end{itemize}

On the other hand, ITU-R Working Party 4B (WP 4B) has carried out studies on performance, availability, air interfaces and earth-station equipment of satellite systems in  Fixed Satellite Services (FSSs) as well as land, maritime, and aeronautical Mobile Satellite Services (MSSs), and Broadcast Satellite Services (BSSs) \cite{ITUWP4B}. This group has given special consideration to the study of \acrfull{ip} related system aspects and performance, and they have developed new and revised recommendations and reports on IP over satellite to meet the increasing demands of satellite links to carry IP traffic. This group cooperates closely with the ITU Telecommunication Standardization Sector.

\paragraph{\textbf{\acrfull{wrc}}} World radio communication conferences are held every three to four years. WRC reviews and, if necessary, revises the Radio Regulations, the international treaty governing the use of the radio-frequency spectrum and the geostationary-satellite and non-geostationary-satellite orbits. Revisions are made on the basis of an agenda determined by the ITU Council, which takes into account recommendations made by previous world radio communication conferences. WRC-19 took place in Sharm El. Sheikh, Egypt on October 28 to November 22, 2019, and 3,540 delegates from 165 countries attended. The members took the following decisions \cite{WRC}:
\begin{itemize}
    \item \acrfull{esim} – expected to provide reliable and high bandwidth internet services to aircraft, ships, and land vehicles.
    \item Resolution lays out technical and regulatory conditions for three types of ESIM communicating with a GSO \acrshort{fss} space stations within the frequency band 17.7-19.7 GHz (space to earth) and 27.5-29.5 GHz (earth to space).
    \item Regulatory frameworks for sharing between GSO and non-GSO satellite systems in the 50/40 GHz range. Also sharing between GSO FSS, \acrshort{bss} \& \acrshort{mss} and non-GSO FSS satellite systems.
   \item Deployment process should be based on  milestone to avoid spectrum warehousing by large non-GSO satellite constellations.
   \item One of the new regulations adopted is that non-GSO systems have to deploy 10\% of the constellation within two years, 50\% within five years and the full deployment within seven years.
\end{itemize}

\paragraph{\textbf{\acrfull{ietf}}} IETF is a large open international community of network designers, operators, vendors, and researchers concerned with the evolution of Internet architecture and the smooth operation of the Internet \cite{IETF}. Under IETF, the Transport and Services Area  covers a range of technical topics related to data transport in the Internet, such as protocol design and maintenance at Layer 4 (e.g., \acrfull{tcp}, \acrfull{udp}, and \acrfull{sctp}), congestion control and (active) queue management, and QoS and related signaling protocols. The transport area subcommittee of the IETF are working with network operators to standardize a  multi-path deployment scheme that includes fixed gateways and satellites for the backhauling of the 5G network. With \acrfull{mptcp},  communication service providers can extend the coverage and the bandwidth of 5G services. This will address the issues where in some areas the fixed network is not able to deliver enough bandwidth for the backhauling. The MPTCP working group has developed mechanisms that add the capability of simultaneously using multiple paths to a regular TCP session without making any assumption about the support of the communicating peers. In this context, satellite networks can be considered as one of the options when more than one path is used for backhauling.

\paragraph{\textbf{\acrfull{etsi}}}
 ETSI has investigated a number of component technologies that will be integrated into future 5G systems, such as \acrfull{nfv}, \acrfull{mec}, \acrfull{mwt}, and \acrfull{nin} \cite{ETSI}. The objective of ETSI activities is to define the end-to-end SatCom system that can be fully integrated in 5G. The standardization work is undertaken by the following technical committees:
\begin{itemize}
    \item Satellite Earth Stations and Systems (TC-SES): Defines all aspects related to satellite earth stations and systems. Within TC-SES, the SCN working group (Satellite Communication and Navigation) covers radio and transmission aspects related to fixed and mobile satellite systems operating in any bands allocated to FSS, MSS, or global navigation satellite systems operating in any bands allocated to RDSS. ETSI TC-SES has established a working relationship with 3GPP. This has allowed the development of standards for mobile satellite as well as broadband satellite multimedia systems, which are mainly based on 3GPP system architecture and radio protocols. 
    \item Network Function Virtualization (ISG-NFV): Produces the technical specifications for the virtualization of network functions.
    \item \acrfull{osm}: Develops a software reference implementation of ETSI \acrfull{mano}.
    \item Multi-Access Edge Computing (TC-MEC): Produces the technical specifications for realizing Multi-Access Edge Computing (MEC) in the context of content delivery (multicasting and caching).
\end{itemize}

\paragraph{\textbf{\acrfull{5gppp}}}
The 5G PPP is a joint initiative between the European Commission and the European ICT industry. The 5G PPP will deliver solutions, architectures, technologies, and standards for the ubiquitous 5G communication infrastructures. In June 2021, The 5G PPP infrastructure association published a new white paper, “Vision and Societal Challenges Work Group," which described the European vision for the 6G network ecosystem \cite{5GPPP}. This white paper covered key areas related to 6G research from a technical, societal, policy, and business perspective, providing a vision for  future mobile networks. The paper dedicated one section to non-terrestrial networks.

\paragraph{\textbf{\acrfull{ecc}}} ECC is an organization of the European Conference of Postal and Telecommunications Administrations. The ECC Working Group Frequency Management (WG FM) is responsible for developing strategies, plans, and implementation advice for the management of the radio spectrum. WG FM44 deals with satellite communications in particular \cite{ECC}.

\paragraph{\textbf{\acrfull{nict} Japan}} The Study Group on the Integration of Satellite Communications and 5G/Beyond 5G of the National Institute of Information and Communications Technology (NICT) Japan identified the use cases that Japan would need by 2040 \cite{japanReport}. The themes of these use cases were the Internet of Things, smart cities, maritime and aviation, transportation infrastructures, and responding to emergency disasters. The use cases  require the integration of satellite networks and \acrshort{5g+} networks. The report produced by NICT highlighted some standardization requirements (e.g., standard communication protocols for every layer) that need to be considered in order to realize the identified use cases. 

\paragraph{\textbf{IEEE \acrfull{ingr} - Satellite Work Group}} IEEE INGR-Satellite WG produced two editions of its roadmap report in 2020 and 2021 \cite{INGR1} \cite{INGR2}. The 2021 Edition of the INGR Satellite Working Group Report discussed several topics related to satellites in future networks, including  applications and services, reference architectures, new MIMO-based PHY, antenna and payload, machine learning and artificial intelligence, edge computing, QoS/\acrfull{qoe}, security, network management and standardization. Under each topic, the related challenges, enablers, and potential solutions were highlighted.

\subsection{Satellite and aviation standardization}

\paragraph{\textbf{Digital Video Broadcasting (DVB)}}  This industry forum was established initially to create a set of standards for the TV broadcasting industry via terrestrial, satellite, or cable networks. DVB has successfully defined the DVB-S2 broadcast channel, which is widely adopted around the world. A Return Channel (DVB-RCS/2) has been defined to support broadband telecommunication services via satellite. The DVB technical specifications were published by ETSI after being reviewed in the TC-BROADCAST. However, DVB technical specifications
have often been associated with some proprietary features at the level of architecture, protocol stack, and radio access. This has resulted in interoperability problems among satellite access networks from different solution vendors, and it has led to a fragmented SatCom market.

\paragraph{\textbf{\acrfull{ccsds}}} CCSDS was formed in 1982 by the major space agencies of the world to provide a forum for discussing common problems in the development and operation of space data systems \cite{CCSDS}. CCSDS has been actively developing standards for data-systems and information-systems to promote interoperability and cross support among cooperating space agencies, to enable multi-agency spaceflight collaboration (both planned and contingency) and new capabilities for future missions. The CCSDS standardization work reduces the cost of spaceflight missions by allowing cost sharing between agencies and cost-effective commercialization.  CCSDS has six technical areas with twenty-three working groups. The working body responsible for defining communications standards is the Space Link Service, composed of six working groups. It defines two main links between earth and space probes: telemetry and telecommand.

\paragraph{\textbf{\acrfull{aeec}}} Develops engineering standards and technical solutions for avionics, networks, and cabin systems that foster increased efficiency and reduced life cycle costs for the aviation community \cite{AEEC}. Their standardization work is overseen by the subcommittee Network Infrastructure and Security, which aims to develop standards for IP connectivity and security for aircraft and to enable fleet-wide solutions based on open standards for lower development costs, increased flexibility, higher reliability, reduced complexity, longer lifespans, and ease of configurability and maintenance. ARINC is AEEC Project Paper 848 - “Secure Broadband IP Air-Ground Interface (SBAGI)," which was intended to define a method for secure communications interface between IP networks contained within an aircraft system and within a ground network hosted by the aircraft original equipment manufacturer, airline or a 3rd party.  The ARINC Project Paper 848  standardized this interface at the network level while taking into account the overall security context.

\section{Standardization for 6G Satellite Communications Networks} \label{sec8}

Most of the standardization work carried out by 3GPP and other standardization organizations focuses on the physical and MAC layers. Consideration has also been given to defining satellite use cases and architectural options in the context of integrated satellite 5G networks. The following subsections highlight several issues that need to be considered in standardization work in order to achieve the complete integration of satellite and terrestrial 6G networks.

\subsection{Mobility management}

LEO satellites provide shorter propagation
delays and higher data rates than GEO satellites. However, these advantages come with the price of frequent handover and topology changes, which yields a time-varying communication channel. Handovers in LEO SatNets are of three types:
\begin{itemize}
    \item Intra-satellite handovers, which occur between satellite beams.
    \item Inter-satellite handovers, which occur between satellites.
    \item Inter-access network handovers (also known as vertical handovers), which occur either between satellites belonging to different access networks or from a SatNet to an ariel network or terrestrial network (or vice versa) in integrated terrestrial-NTN systems.
\end{itemize}

In 6G future networks, LEO SatNets will not only serve rural or remote areas but will also provide communication services and coverage in urban and highly populated areas. Such a scenario will lead to thousands of UE being connected to an LEO satellite and this large group of users will need to go through a frequent handover process at almost the same time. Managing the handover of thousands of users simultaneously or semi-simultaneously using conventional handover management schemes will create huge network loads. New handover management schemes are required to deal with this issue in 6G LEO SatNets. 

For mobility management in IP-based networks, IETF introduced a number of protocols, such as \acrfull{mip} and \acrfull{pmip}. However, such protocols were not designed to deal with the high topology change rate in SatNets, where everything is moving including the gNB (LEO satellite base station). A number of approaches have been proposed to address this problem \cite{darwish2021location}. Nevertheless, the concept of separating control plane and data plane of \acrfull{sdn} is a promising approach to efficiently manage SatNet topology. 

The fast-moving footprint of LEO satellites affects the paging procedure, which is primarily related to the tracking area management. The tracking area is the satellite coverage area (footprint); it can be fixed or moving. Although the moving tracking area accommodates the LEO satellite moving footprint, it results in high paging loads that are difficult to manage by the network. In addition, supporting dual-connectivity and
 vertical handovers in future LEO SatNets requires novel mechanisms to provide seamless mobility in integrated 6G networks and to improve global network coverage and service.

\subsection{Routing}
One very important characteristic of LEO mega-constellations is the ability of satellites to form networks and communicate with each other through ISLs \cite{9351765}. Due to the frequent topology changes in an LEO SatNet, ISLs have a limited lifetime. In addition, some ISLs may get congested due to high traffic loads at certain partitions of the SatNet. Moreover, as LEO SatNets are expected to serve different types of applications, there are certain QoS requirements (e.g., packet delivery delay, packet delivery ratio) that need to be met for each type of applications. Therefore, successful data delivery will require robust routing schemes that can fulfil the QoS requirements of each application type and adapt to the unique characteristics of LEO SatNets. For example, delay-tolerant routing is suitable for delay-tolerant applications, while multi-path routing is required to support applications with high bandwidth requirements. Thus, it is crucial to develop standard routing protocols that adapt to the SatNet dynamic environment and satisfy the various user application requirements. Standards should support interoperability among the different satellite constellations and operators. Moreover, cross network routing (i.e., across satellite, aerial, and terrestrial networks) should be considered to achieve the full integration in 6G. To support efficient routing, topics such as resource allocation, network monitoring, and congestion control should be considered as part of the standardization work.

\subsection{Adoption of SDN/NFV}
The SDN/NFV paradigms will play a key role in future integrated networks. However, the use of SDN/NFV in an LEO SatNet has not yet been fully investigated. Several software-defined satellite network architectures have been proposed in the literature, such as centralized, distributed, and cluster-based \cite{darwish2021location}. Nevertheless, SDN-based solutions for SatNets should be considered in  standardization works to provide compatibility and interoperability among the integrated network components and the different vendors and service providers. For instance, on-board SDN-compatible routers could be developed following a specific standard to operate on LEO satellites and provide a softwareized routing function that can adapt to changes in the dynamic environment of LEO SatNets. 

NFV is going to be particularly necessary to hide the complications of integrated networks from the user side. NFV can be used in various applications, such as the virtualization of mobile base stations, content delivery networks, and platforms as a service. The virtualization of network functions deployed on general purpose standardized hardware is expected to reduce service and product introduction times as well as capital and operational expenditures. According to ETSI, an important part of  the NFV environment control  should be accomplished through automation and orchestration. ETSI created a separate stream, MANO, within NFV describing how flexibility should be controlled. ETSI introduced a full set of standards to enable an open ecosystem where \acrfull{vnf} can be interoperable with independently developed management and orchestration systems. Many major network equipment vendors have announced support for NFV. On the other hand, major software suppliers announced that they will be providing NFV platforms to be used by equipment suppliers to build their NFV products. However, in the area of satellite networks, the adoption of these concepts and technologies is still in its infancy. Further investigations are required to identify the requirements needed to adopt NFV in SatNets. In addition, the support for NFV should be considered in the design of satellite network components. 

\subsection{Intelligent management and orchestration} 
AI and ML will be an integral part of 6G networks, especially at the level of network management and orchestration. ETSI launched the \acrfull{isg} on \acrfull{eni} in February 2017 \cite{ETSIENI}. ENI is an entity that provides intelligent network operation and management recommendations and/or commands to an assisted system (i.e., an existing system leveraging the intelligent capabilities of ENI). ENI has two operation modes: Recommendation
mode and Management mode. The former provides insights and advice to the operator or assisted system, whereas the latter may also provide policy
commands to the assisted system \cite{wang2020design}. In another effort to advance network automation, 3GPP introduced the concept of SON \cite{3gpp32500} where AI/ML can be applied to automate several network management functions. However, both ENI and SON concepts are still limited to the 5G context and may not be sufficiently agile in coping with the immense levels of complexity, heterogeneity, and mobility in the envisioned beyond-5G integrated networks. To support the intelligence and autonomous nature of 6G, the concept of \acrfull{sen} was presented in \cite{darwish2020vision} \cite{farajzadeh2021self}. SEN considers the integrated architecture of 6G and beyond. SEN utilizes AI/ML to make future integrated networks fully automated, and it intelligently evolves with respect to the provision, adaptation, optimization, and management aspects of networking, communications, computation, and infrastructure nodes' mobility. SEN can be adopted to support real-time decisions, seamless control, intelligent management in SatNets  to achieve high-level autonomous operations. Nevertheless, SEN is quite a recent concept and has not yet been considered by standardization organizations. 

\subsection{Dynamic spectrum management}

Dynamic and efficient spectrum management is important in SatNets due to the pervasive growth of wireless communications and the ever-increasing demands by bandwidth-hungry UE \cite{liang2021realizing}. The problem of spectrum scarcity is one of the key challenges facing future SatNets as  more satellites are deployed and more applications are emerging. The factors of unpredictable user mobility and satellite mobility make dynamic spectrum allocation necessary but difficult as well. Dynamic spectrum allocation needs to be considered on multiple levels to mitigate inter-cell interference in multi-beam satellite systems, inter-satellite interference, and interference between satellites and terrestrial communications when bands are shared. In addition, spectrum management must consider higher frequency bands (THz) \cite{tekbiyik2020reconfigurable} and the option of using free space optical (FSO) \cite{chaudhry2020free}   communications as they are expected to be utilized in future SatNets. Although various kinds of static and dynamic spectrum allocation schemes have been studied by satellite researchers, this issue is not covered sufficiently in the standardization works.

\section{Conclusion}\label{sec9}

To merge the ecosystems of satellite and terrestrial communications and steer the research and standardization communities from classical bent-pipe satellite systems to the mega-constellation satellite systems is not an easy task. Standardization is one of the main enablers for the integration of satellites and terrestrial networks. The 3GPP community has achieved some advancements in the NTN integration with 5G from a standardization perspective. However, more standardization efforts are needed to realize the full integration of SatNets and 5G+ on the physical layer level up to the application level.

\ifCLASSOPTIONcompsoc
  \section*{Acknowledgments}
\else
  \section*{Acknowledgment}
\fi

This work is supported in part by MDA Space, and in part by Mitacs Canada.

\bibliographystyle{IEEEtran}
\bibliography{IEEEabrv,ref}




\begin{IEEEbiography}[{\includegraphics[width=1in,height=1.25in,clip,keepaspectratio]{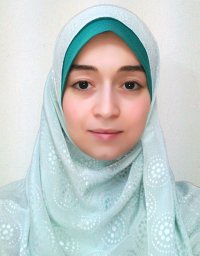}}]{Tasneem Darwish} is currently a postdoctoral fellow at the Department of Systems and Computer Engineering, Carleton University, Canada.  She received the MSc. degree with merit in Electronics and Electrical Engineering from the University of Glasgow, UK, in 2007 and her Ph.D. degree in Computer Science from Universiti Teknologi Malaysia (UTM), Malaysia, in 2017. From  2017 to 2019, Tasneem was a postdoctoral fellow at UTM . From 2019 to 2020, she was a research associate at Carleton University, Canada. In 2020, Tasneem started working as a postdoctoral fellow at Carleton University on a collaborative project with MDA Space to investigate mobility management in future LEO satellite networks. She is the recipient of the UTM Alumni Award for Science and Engineering in 2017. She was awarded the Malaysia International Scholarship (MIS) from 2013 to 2016. Her current research interests include mobility management in future LEO satellite networks, edge/fog computing and data offloading in HAPS, vehicular ad hoc networks, and intelligent transportation systems.  Tasneem is a senior IEEE member and an active reviewer for several IEEE journals such as IEEE Internet of Things, IEEE Access, IEEE Transactions on Vehicular Technology, and IEEE Transactions on Intelligent Transportation Systems. 
\end{IEEEbiography}

\begin{IEEEbiography}[{\includegraphics[width=1in,height=1.3in]{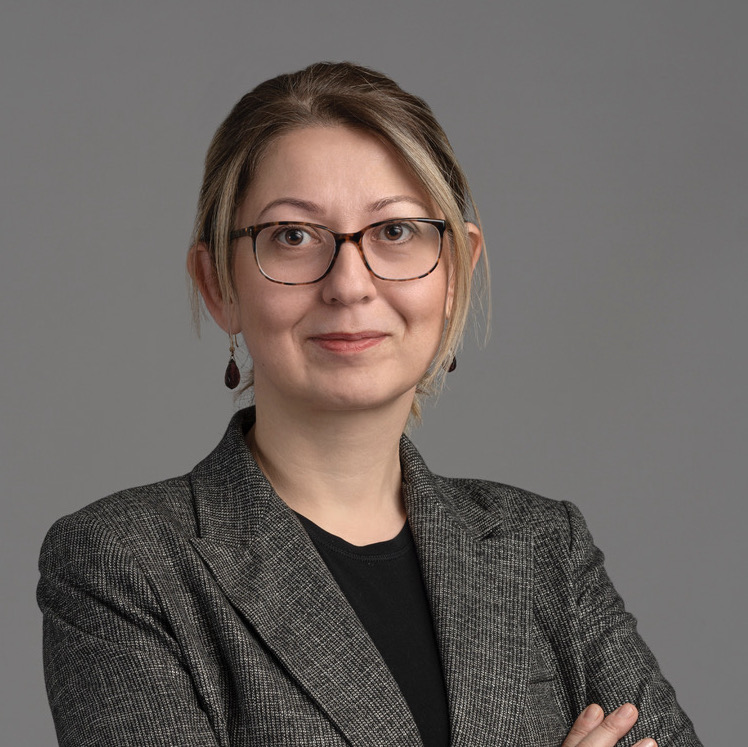}}]{Gunes Karabulut Kurt} is currently an Associate Professor of Electrical Engineering at Polytechnique Montréal, Montreal, QC, Canada.  She received the B.S. degree with high honors in electronics and electrical engineering from the Bogazici University, Istanbul, Turkey, in 2000 and the M.A.Sc. and the Ph.D. degrees in electrical engineering from the University of Ottawa, ON, Canada, in 2002 and 2006, respectively. From 2000 to 2005, she was a Research Assistant at the University of Ottawa. Between 2005 and 2006, Gunes was with TenXc Wireless, Canada. From 2006 to 2008, she was with Edgewater Computer Systems Inc., Canada. From 2008 to 2010, she was with Turkcell Research and Development Applied Research and Technology, Istanbul. Gunes has been with Istanbul Technical University since 2010, where she is currently on a leave of absence. She is a Marie Curie Fellow and has received the Turkish Academy of Sciences Outstanding Young Scientist (TÜBA-GEBIP) Award in 2019. She is an Adjunct Research Professor at Carleton University. She is also currently serving as an Associate Technical Editor (ATE) of the IEEE Communications Magazine and a member of the IEEE WCNC Steering Board.
\end{IEEEbiography}

\begin{IEEEbiography}[{\includegraphics[width=1in,height=1.3in]{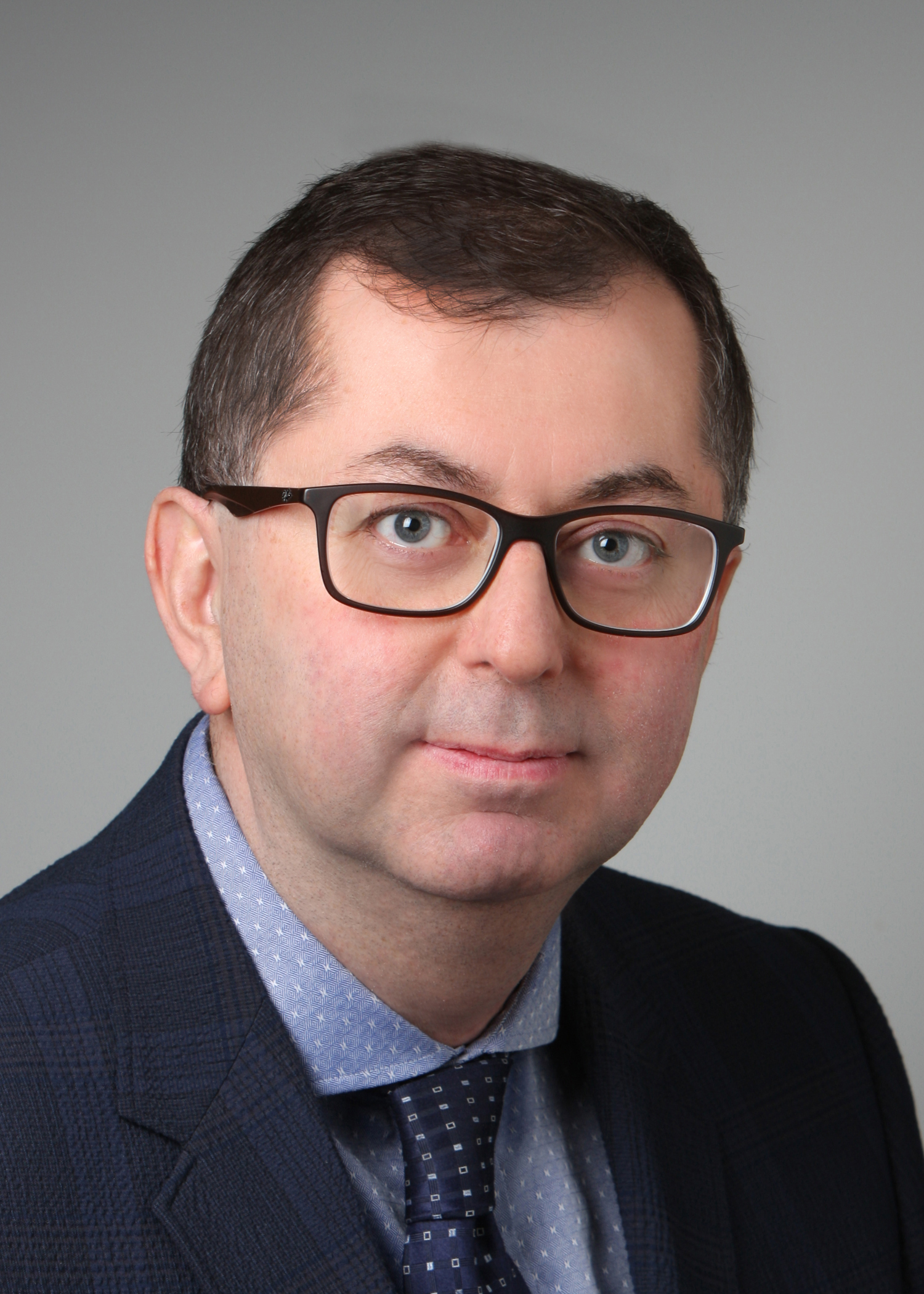}}]{Dr. Halim Yanikomeroglu} is a Professor in the Department of Systems and Computer Engineering at Carleton University, Ottawa, Canada. His primary research domain is wireless communications and networks. His research group has made substantial contributions to 4G and 5G wireless technologies. His group’s current focus is the aerial (UAV and HAPS) and satellite networks for the 6G and beyond-6G era. His extensive collaboration with industry resulted in 37 granted patents. He is a Fellow of IEEE, EIC (Engineering Institute of Canada), and CAE (Canadian Academy of Engineering), and a Distinguished Speaker for both IEEE Communications Society and IEEE Vehicular Technology Society. He is currently serving as the Chair of the IEEE WCNC (Wireless Communications and Networking Conference) Steering Committee. He served as the General Chair or TP Chair of several conferences including three WCNCs and two VTCs. He also served as the Chair of the IEEE’s Technical Committee on Personal Communications. Dr. Yanikomeroglu received several awards for his research, teaching, and service, including the IEEE ComSoc Fred W. Ellersick Prize in 2021, IEEE VTS Stuart Meyer Memorial Award in 2020, and IEEE ComSoc Wireless Communications Technical Committee Recognition Award in 2018.
\end{IEEEbiography}

\begin{IEEEbiography}[{\includegraphics[width=1in,height=1.3in]{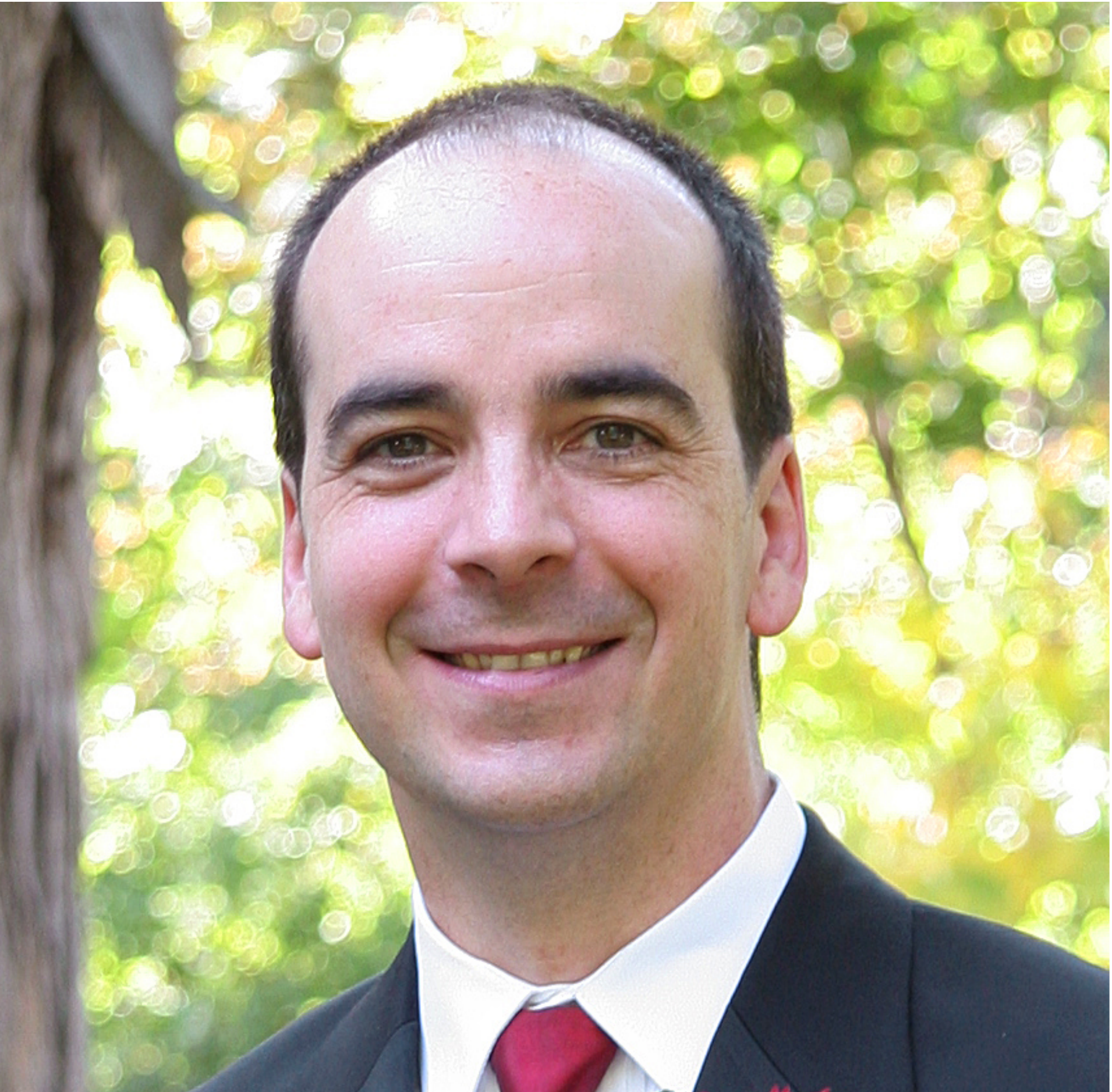}}]{Michel Bellemare}received is B.Eng. degree from Université de Sherbrooke, Sherbrooke, Québec, Canada in Communications Engineering in 1986. He has gained experience in terrestrial and space wireless communications in various companies such as Nortel Networks, Ultra Electronics, SR Telecom, etc. He is now a Space Systems Architect at MDA corporation, Sainte-Anne-de-Bellevue, Québec, Canada.
\end{IEEEbiography}

\begin{IEEEbiography}[{\includegraphics[width=1in,height=1.3in]{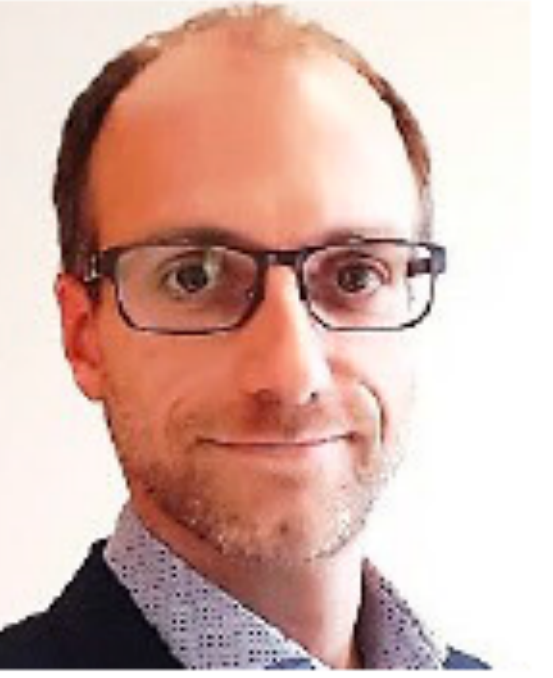}}]{Guillaume Lamontagne} received the B.Eng. and M.Eng. degrees from the École de Technologie Supérieure (ÉTS), Montréal, QC, Canada, in 2007 and 2009 respectively. His experience in satellite communications started through internships and research activities with the Canadian Space Agency (CSA), in 2005, and the Centre national d’études spatiales (Cnes), France, in 2006 and 2008. He joined MDA in 2009 and held various communication systems engineering and management positions before being appointed as the Director of Technology, Payloads, in 2019. Through this role, he is leading MDA’s Research and Development activities for satellite communications as well as establishing the related long term development strategy.
\end{IEEEbiography}

\end{document}